\documentclass{JHEP3}
\usepackage{epsfig}

\newcommand{\p}{\partial}
\newcommand{\nn}{\nonumber}
\newcommand{\nnn}{\nonumber \\}
\newcommand{\ti}{\tilde}
\newcommand{\mt}{\mathcal}

\preprint{HUPD0806}

\title{Electroweak Symmetry Breaking and Singlino Dark Matter with Deflected Anomaly Mediation}

\author{Norimi Yokozaki \\

    Graduate School of Science, Hiroshima University, \\
    Higashi-Hiroshima, Hiroshima 739-8526, Japan. \\
    E-mail: \email{norimi@theo.phys.sci.hiroshima-u.ac.jp}
}

\abstract{
We investigate the phenomenology of the Nearly Minimal Supersymmetric Standard Model (nMSSM) in the deflected anomaly mediation scenario. 
We also include the Fayet-Iliopoulos D-term of the standard model gauge group.
In the nMSSM, the mu term is replaced by the vacuum expectation value of the gauge singlet;
therefore, there is no difficulty in generating the B-term of the SUSY breaking scale. 
Although the messenger sector is introduced, direct couplings between 
nMSSM fields 
and messenger sector fields are forbidden by the discrete symmetry. 
Therefore, the phenomenology at the weak scale does not 
depend on the detail of the messenger sector. 
We show that there are regions of parameter space in which electroweak symmetry breaking occurs successfully
 and the lightest Higgs is heavier than the LEP bound.
We show that the gluino is light in this scenario.
The lightest neutralino, which is mainly composed of a singlino, is a candidate for dark matter. 
The relic density explains the observed abundance of dark matter.
The dark matter-nucleon scattering cross section satisfies the current limit from CDMS and XENON10
with a small value for the strange quark content of a nucleon.
}


\begin{document}

\section{Introduction}
The Minimal Supersymmetric Standard Model (MSSM) is the most attractive framework for the physics beyond the Standard Model.
In the MSSM, gauge coupling unification is achieved and the Higgs potential is stabilized. 
Despite these good features, the MSSM has difficulty in the Higgs sector. 
The MSSM has a $\mu$ term, $\mu H_1 H_2$, in the superpotential. To 
maintain the weak-scale vacuum expectation value (VEV) of a Higgs, 
$|\mu|$ has to be at the weak scale. However, it is difficult to explain why such a dimensionful parameter
is much smaller than Plank scale or GUT scale. This problem is the so-called $\mu$ problem.

A simple way of solving the $\mu$ problem is to introduce a gauge singlet, and replace the $\mu$ by the VEV of the gauge singlet field:
\begin{eqnarray}
\mu H_1 H_2 \rightarrow \lambda \left<S\right> H_1 \cdot H_2 \ .
\end{eqnarray}
The most famous model to include a gauge singlet is the Next-to-Minimal Supersymmetric Standard Model (NMSSM).
In the NMSSM, a new discrete symmetry, $Z_3$, is introduced to forbid the mass term for $S$.
However, $Z_3$ symmetry spontaneously breaks down when electroweak symmetry breaking occurs. 
At this point, unacceptably large cosmological domain walls appear\cite{domainwall}.
In the Nearly Minimal Supersymmetric Standard 
Model (nMSSM)\cite{nmssm1}\cite{nmssm2}\cite{nmssm3}, the cosmological 
domain wall problem is solved by tadpoles. The tadpoles are generated by 
supergravity interactions and explicitly break the discrete symmetry. Therefore, the 
domain wall problem does not arise.
The nMSSM has the same attractive feature of electroweak baryogenesis as 
the NMSSM has. To achieve successful electroweak baryogenesis,
a strong first-order phase transition is required. Therefore, new sources of CP-violation beyond the CKM matrix have to exist. 
In the nMSSM, there are additional sources of CP-violation in the singlet sector. Therefore, unlike MSSM, 
nMSSM does not rely on radiative contributions from a light stop for strong first-order phase transition
\cite{nmssm_ewbg1}\cite{nmssm_ewbg2}.

SUSY breaking terms are important in discussing phenomenology, SUSY breaking effects are transmitted to the nMSSM sector from
a hidden sector by one or more mediation schemes.
One interesting mediation scheme is anomaly mediation \cite{am-lisa}\cite{am-org}\cite{am-sing}.
In anomaly mediation, the supergravity actions of the hidden sector and visible sector are sequestered.
SUSY breaking effects are transmitted to the visible sector due to the superconformal anomaly. There are studies in which
soft breaking terms are derived by anomaly mediation with NMSSM-like models\cite{am-nmssm}\cite{am-fat}.
In these works, successful electroweak symmetry-breaking is achieved; however, the VEV of $S$ is on the order of a few TeV. 
This leads to a large higgsino mass: $\mu_{eff} = \lambda \left<S\right>$.
Therefore, there are large mass splittings among Higgs (and neutralinos).

To obtain a moderate value for $\mu_{eff}$, we consider a deflected anomaly mediation scenario \cite{dam-org}\cite{dam-ph}\cite{dam-pos}, 
which introduces an additional messenger sector. The SUSY breaking mass for a messenger is given by a VEV of the gauge singlet field, $X$.
In the original deflected anomaly mediation scenario\cite{dam-org}\cite{dam-ph},
the superpotential of $X$ is extremely flat; therefore, the fermionic 
component of $X$, $\psi_X$, becomes light and the lightest SUSY particle is $\psi_X$. 
In the positively deflected anomaly mediation scenario\cite{dam-pos}, the superpotential is not flat; therefore, $\psi_X$ does not have to
be light\cite{dam-recent} and an ordinary SUSY particle can be a candidate for dark matter.
We consider the positively deflected anomaly mediation scenario.
We also consider SUSY breaking with the Fayet-Iliopoulos D-term. 

We show that when nMSSM and deflected anomaly mediation are combined,
successful electroweak symmetry breaking occurs for a moderate value of $\mu_{eff}$.
We also show that the lightest neutralino, which is mainly composed of a 
singlino, is a good candidate for dark matter. We also present 
sparticle mass spectra. 

This paper is organized as follows. In section 2, we introduce the nMSSM Lagrangian and discuss tadpoles. We also discuss the
 direct couplings between nMSSM fields and messenger sector fields.
In section 3, we derive the soft SUSY breaking terms of
the nMSSM fields in the deflected anomaly mediation scenario.
Section 4 is devoted to the phenomenology of
this scenario. Finally, section 5 presents our conclusions.

\section{Nearly Minimal Supersymmetric Standard Model}
In this section, we discuss tadpoles and direct couplings between 
nMSSM fields and messenger sector fields. First, we introduce the nMSSM
Lagrangian.

The superpotential and soft breaking terms in the nMSSM are 
\begin{eqnarray}
W_{nMSSM} &=& \lambda \hat{S} \hat{H_1} \cdot \hat{H_2} + \frac{m_{12}^2}{\lambda}\hat{S} + 
y_u \hat{Q} \cdot \hat{H_2} \hat{U}^c +
y_d \hat{Q} \cdot \hat{H_1} \hat{D}^c \nnn
&& + y_l \hat{L} \cdot \hat{H_1} \hat{E}^c \label{eq:nmssm_sp} \ ,
\end{eqnarray}
and
\begin{eqnarray}
-\mathcal{L}_{soft} &=& m_S^2 |S|^2 + (a_\lambda S H_1 \cdot H_2 + h.c.) + (t_S S + h.c.) \nonumber \\
&& + \tilde{m}_{H_1}^2 H_1^\dagger H_1 + \tilde{m}_{H_2}^2 H_2^\dagger H_2 \nonumber \\
&& + \tilde{m}_Q^2 \tilde{Q}^\dagger \tilde{Q} + \tilde{m}_U^2 |\tilde{u_R}|^2 + \tilde{m}_D^2 |\tilde{d_R}|^2 
+ \tilde{m}_L^2 \tilde{L}^\dagger \tilde{L} + \tilde{m}_E^2 |\tilde{e_R}|^2 \nonumber \\
&& + (a_u \tilde{Q} \cdot H_2 \tilde{u_R}^* + a_d \tilde{Q} \cdot H_1 
 \tilde{d_R}^* + a_l \tilde{L} \cdot H_1 \tilde{e_R}^* + h.c.) \ . \label{eq:nmssm_soft}
\end{eqnarray}
$\hat{S}$ denotes a gauge singlet chiral superfield, and $S$ is the scalar component of $\hat{S}$. 
When $S$ acquires the VEV, the higgsino mass parameter, $\mu_{eff} = \lambda v_s$, is
generated effectively. We take $\lambda$ to be real positive by suitable redefinitions of $S$, $H_1$ and $H_2$.
Unlike the NMSSM, there are 
no trilinear terms for the gauge singlet.
$m_{12}^2 \hat{S}/\lambda$ and $t_S S$ are tadpoles. They are absent at the tree level; 
however, they are generated radiatively by supergravity interactions. 
These terms are on the order of the weak scale, as we describe below.

\subsection{Tadpoles}
The greatest difference between the nMSSM and NMSSM is the existence of 
tadpoles in the former. The tadpoles are generated by supergravity interaction.
In the nMSSM, the theory has a global discrete symmetry at tree level.
This symmetry guarantees that the generated tadpoles are on the order of the weak scale,
despite the fact that supergravity interactions break global symmetries.
Because the tadpoles explicitly break the discrete symmetry,
the domain wall problem does not appears.

In nMSSM, the Lagrangian has a discrete R symmetry $Z_{nR'}$ at tree 
level. The charge assignment of the fields is shown in 
Table \ref{table:nmssm_symmetry}.
The charge of $Z_{nR'}$, $Q_{nR'}$, is defined as
\begin{eqnarray}
Q_{PQ} + 3 Q_R \ , 
\end{eqnarray}
where $Q_{PQ}$ denotes the charge of Peccei-Quinn symmetry, 
$U(1)_{PQ}$, and $Q_R$ denotes the charge of $U(1)_R$.
Under $Z_{nR'}$, the nMSSM fields transform as
\begin{eqnarray}
 \Phi_i \rightarrow \Phi_i \exp\left({i \frac{Q_{nR'}}{n} \theta}\right),
\end{eqnarray}
where $\Phi_i$ denotes nMSSM fields. If the Lagrangian respects $Z_{5R'}$ 
or $Z_{7R'}$ at the tree level, the scale of the generated tadpoles 
can naturally be the weak scale\cite{nmssm1}\cite{nmssm3}.
When the discrete symmetry is $Z_{5R'}$, tadpoles of six-loop order are generated.
 When the discrete symmetry is $Z_{7R'}$, tadpoles of seven-loop order are generated.
The tadpoles break the $Z_{5R'}$ or $Z_{7R'}$, and therefore no cosmological domain wall problem exists.

\TABLE[thbp]{
\begin{tabular}{|c|c|c|c|c|c|c|c|c|c|}
\hline
& $\hat{H}_1$ & $\hat{H}_2$ & $\hat{S}$ & $\hat{Q}$ & $\hat{L}$ & $\hat{U}^c$ & $\hat{D}^c$ & $\hat{E}^c$ & $W$ \\
\hline
$U(1)_{PQ}$ & 1 & 1 & -2 & -1 & -1 & 0 & 0 & 0 & 0 \\
\hline
$U(1)_R$ & 0 & 0 & 2 & 1 & 1 & 1 & 1 & 1 & 2 \\
\hline
$Z_{nR'}$& 1 & 1 & 4 & 2 & 2 & 3 & 3 & 3 & 6 \\ 
\hline
\end{tabular}
\caption{Charge assignments of fields}
\label{table:nmssm_symmetry}
}

The generated tadpoles are given by\cite{nmssm3}:
\begin{eqnarray}
V_{tad} \sim \frac{1}{(16\pi^2)^l} \left(M_p M_{susy}^2 S +M_{susy} F_S + h.c. \right),
\end{eqnarray}
where $l$ is the number of the loops at which tadpoles first appear. 
$l=6$ in the $Z_{5R'}$ case and $l=7$ in $Z_{7R'}$ case.
In a deflected anomaly mediation scenario as well as an anomaly mediation scenario,
$M_{susy}$ is $\mt{O}(10 {\rm 
TeV})$; therefore, $Z_{7R'}$ is favorable. 
As we describe below, $Z_{7R'}$ also forbids direct couplings between 
nMSSM fields and messenger sector fields.

\subsection{Direct Couplings to the Messenger sector}
In a deflected anomaly mediation scenario, the messenger sector is 
introduced in addition to the hidden sector, which is the origin of SUSY breaking.
The messenger sector contains a gauge singlet chiral superfield and messenger superfields.
The messengers transmit the SUSY breaking to the nMSSM sector, and this SUSY breaking is 
comparable to that of anomaly mediation. 
In this subsection, we show that direct couplings between the messenger sector fields and the nMSSM fields do not exist.

We consider the following superpotential in the messenger sector.
\begin{eqnarray}
W_{mess} = \frac{1}{2} m_X \hat{X}^2 + \lambda_X \hat{X} \bar{\Psi}_i{\Psi^i} \ ,\label{eq:mess}
\end{eqnarray}
where $\hat{X}$ is a gauge singlet chiral superfield. $\bar{\Psi}_i$ and ${\Psi^i}$ are the messenger fields 
that transform ${\bf \bar{5}}$ and ${\bf 5}$ for the 
$SU(5)$ GUT gauge group respectively. 
The $Z_{nR'}$ charge assignment of the fields is shown in Table \ref{table:nmssm_gm_charge}.

\TABLE[thbp]{
\begin{tabular}{|c|c|c|c|c|}
\hline
& $X$ & $\Psi$ & $\bar{\Psi}$ & $W_{mess}$ \\
\hline
$U(1)_{PQ}$ & 0 & 0 & 0 & 0\\
\hline
$U(1)_R$ & 1 & 1/2 & 1/2 & 2\\
\hline
$Z_{nR'}$& 3 & 3/2 & 3/2 & 6\\
\hline
\end{tabular}
\caption{Charge assignment of the fields in the messenger sector}
\label{table:nmssm_gm_charge}
}

In this charge assignment, there are no direct couplings between messenger sector fields and nMSSM fields.
A direct coupling between the messengers and nMSSM gauge singlet, $S \bar{\Psi}_i{\Psi}^i$, is forbidden by $Z_{nR'}$ symmetry.
$S X^2$ and $X H_u H_d$ terms are also forbidden. On the other hand, the $S^2 X$ term is forbidden by $Z_{7R'}$ but allowed by $Z_{5R'}$. 
In the discussion about tadpoles in the previous subsection, we assumed 
that $Z_{7R'}$ symmetry exists at tree level.
Therefore, there are no direct couplings between nMSSM fields and the messenger sector fields.
The phenomenology of the nMSSM at the weak scale does not depend on the detail of messenger sector.

\section{SUSY Breaking}
In this section, we derive the SUSY breaking terms of the nMSSM in the deflected anomaly 
mediation scenario. We also show the corrections to the soft scalar mass 
with the Fayet-Iliopoulos D-term.

In the original deflected anomaly mediation scenario,
 the superpotential of the gauge singlet $\hat{X}$ is flat; therefore in general,
the lightest SUSY particle (LSP) is the fermionic component of $X$, $\psi_X$.
 Threshold corrections to the sparticle mass squared are negative. 
In the positively deflected anomaly mediation scenario,
the superpotential of $\hat{X}$ is not flat; 
therefore, $\psi_X$ is not necessarily the LSP. Corrections to the sparticle mass squared are positive. 
We consider the positively deflected anomaly mediation scenario.
In Appendix A, we give an 
explicit example of the positively deflected anomaly mediation scenario 
in which the fermionic partner of $X$ is not the LSP.

When $\hat{X}$ acquires the VEV,
messengers obtain SUSY breaking mass, $X + F_X \theta^2$. This SUSY breaking mass introduces an intermediate threshold 
that depends on $\theta^2$.
In deflected anomaly mediation, 
corrections from anomaly mediation to soft breaking terms are generated by the following threshold.
\begin{eqnarray}
\frac{X+F_X \theta^2}{\Lambda \hat{\phi}} = \frac{X}{\Lambda} \left[1+\left(\frac{F_X}{X}-F_\phi\right)\theta^2 \right] \equiv \frac{X}{\Lambda}\left(1+dF_\phi \theta^2\right) , \label{eq:thre}
\end{eqnarray}
where $\hat{\phi}$ is the chiral compensator field, $\hat{\phi}=1+F_\phi 
\theta^2$, and $\Lambda$ is the ultraviolet cutoff. $d$ is the deflection 
parameter, which denotes the threshold correction to the SUSY breaking. 
In the positively deflected anomaly mediation scenario, $d$ is positive.


Messengers also affect the beta-functions of gauge couplings. The 
beta-functions of gauge couplings above the scale $|X|$ are written as
\begin{eqnarray}
\frac{d g_a}{d \ln \mu} = -\frac{g_a^3}{16\pi^2} (b_a -N_f) ,
\end{eqnarray}
where $N_f$ is the number of messengers. 
For the intermediate threshold and modification of the beta-functions, 
soft breaking terms in an anomaly mediation are changed to 
those of a deflected anomaly mediation scenario.

In a deflected anomaly mediation, the gaugino mass, soft breaking mass and 
scalar trilinear couplings at the scale $\mu$ are obtained using the following relations\cite{dam-org}\cite{dam-pos}.
\begin{eqnarray}
\frac{m_\lambda(\mu)}{g^2(\mu)} &=& -\frac{F_\phi}{2}\left(\frac{\p}{\p \ln\mu} -d\frac{\p}{\p\ln |X|}\right) g^{-2}\left(\frac{\mu}{\Lambda},\frac{|X|}{\Lambda}\right) \nnn
\tilde{m}_i^2(\mu) &=& -\frac{|F_\phi^2|}{4}\left(\frac{\p}{\p\ln\mu}-d\frac{\p}{\p\ln|X|}\right)^2 \ln Z_i \left(\frac{\mu}{\Lambda},\frac{|X|}{\Lambda}\right) \nnn
\frac{a_{ijk}(\mu)}{y_{ijk}(\mu)} &=& - \frac{F_\phi}{2}\left(\frac{\p}{\p\ln\mu}-d\frac{\p}{\p\ln|X|}\right) \sum_{l=i,j,k} \ln Z_l \left(\frac{\mu}{\Lambda},\frac{|X|}{\Lambda}\right), \nnn
\end{eqnarray}
where $F_\phi$ is the F-term of the chiral compensator fields and corresponds to the gravitino mass. $|X|$ is the messenger scale and $\mu < |X|$. $d$ 
is defined in eq. (\ref{eq:thre}).
$|X|$ and $d$ can be determined by the superpotential and the soft breaking terms in 
the messenger sector (see Appendix A). However, we treat them as the parameters of SUSY breaking because we focus on the 
phenomenology at the weak scale.

In general, the formula for soft breaking terms is complicated. However, by setting the scale as $\mu = |X|$,
 the soft breaking terms are simplified as
\begin{eqnarray}
m_{\lambda_a} &=& -\frac{g_a^2}{(4\pi)^2}\left(b_a - dN_f\right) F_\phi 
 \ ,\nnn
\tilde{m}_i^2 &=& \frac{|F_\phi|^2}{2(4\pi)^4} \sum_a c_a^i g_a^4 \left[b_a + d(d+2)N_f\right] \nnn
&& - \frac{|F_\phi|^2}{4} \sum_y \frac{\p \gamma_i(|X|)}{\p y} \beta_y(|X|) \ ,\nnn
a_{ijk} &=& -\frac{F_\phi}{2} \left[\gamma_i(|X|) + \gamma_j(|X|) + \gamma_k(|X|) \right] y_{ijk} \label{eq:formula}.
\end{eqnarray}
Here, $N_f$ is the number of messengers. $b_a$ and $c_a^i$ are the coefficients 
of the gauge coupling beta functions and the anomalous dimensions of the 
fields respectively.
$b_a = (-33/5, -1, 3)$, $c_a^L=(3/5, 3, 0)$, $c_a^{E^c}=(12/5, 0, 0)$, $c_a^Q=(1/15, 3, 16/3)$,
$c_a^{U^c}=(16/15, 0,16/3)$ and $c_a^{D^c}=(4/15, 0, 16/3)$.

The formula for gaugino masses is easily obtained with
\begin{eqnarray}
g_a^{-2}(\mu) = g_a^{-2}(\Lambda) + \frac{b_a}{8\pi^2} \ln\frac{\mu}{|X|} + \frac{b_a-N_f}{8\pi^2}\ln\frac{|X|}{\Lambda}. \label{eq:gg}
\end{eqnarray}
Equation (\ref{eq:gg}) can be obtained by integrating the beta-functions explicitly.
The derivations of $\ti{m}_i^2$ and $a_{ijk}$ are given in Appendix B. 

For the first and second generations of squarks and sleptons, we can
neglect the contributions from Yukawa couplings. However, for the soft scalar 
mass and the A-term of the third generation of squarks and sleptons,
 the contributions from Yukawa couplings are important. 
For $\tilde{m}_S^2$, $\tilde{m}_{H_1}^2$, $\tilde{m}_{H_2}^2$ and $a_\lambda$, contributions from
Yukawa couplings are also important. The anomalous dimensions of $H_1$ and $H_2$ are different from those of the MSSM due to $\lambda$
and are given in Appendix C. 
The anomalous dimensions of the other fields are same as those of the MSSM and are given in \cite{beta-functions}.
When Yukawa couplings are small,
we obtain the results of \cite{dam-org}\cite{dam-pos}.

In a supersymmetric model, there is an additional source of SUSY breaking, the Fayet-Iliopoulos D-term. 
This term contributes to the square of the scalar mass.

The Fayet-Iliopoulos D-term is
\begin{eqnarray}
\mathcal{L} \ni - \xi D .
\end{eqnarray}
The D-term of the Lagrangian is written as
\begin{eqnarray}
\mathcal{L}_D = \frac{1}{2} D^2 - g D \sum_i q_i A_i^\dagger A_i - \xi D , \label{eq:dterm-xi}
\end{eqnarray}
where $q_i$ is the U(1) charge of the field $A_i$. After eliminating the 
D-term with the equation of motion,
the Lagrangian $\mathcal{L}_D$ becomes
\begin{eqnarray}
\mathcal{L}_D = - \frac{1}{2} \left( \sum_i q_i A_i^\dagger A_i + \xi 
			     \right)^2 \ .
\end{eqnarray}
This leads to additional contributions to the scalar mass terms:
\begin{eqnarray}
\ti{m}_{ij}^2 \rightarrow \ti{m}_{ij}^2 + q_i \xi \delta_{ij} .
\end{eqnarray}

In the Supersymmetric Standard Model, there is only one $U(1)$ gauge group.
In the lepton sector, the hypercharge of the $SU(2)$ doublet is $-1$ and the 
hypercharge of the $SU(2)$ singlet is $+2$. Therefore, we can not solve the tachyonic slepton mass problem 
in anomaly mediation using only the $U(1)_Y$ D-term. 

So far, the additional contributions from the D-term to the soft 
breaking mass of the nMSSM matter fields
are
\begin{eqnarray}
&& \delta \tilde{m}_L^2 = -D_Y \frac{|F_\phi|^2}{(4\pi^4)} \ ,\nnn
&& \delta \tilde{m}_E^2 = 2D_Y \frac{|F_\phi|^2}{(4\pi^4)} \ ,\nnn
&& \delta \tilde{m}_Q^2 = \frac{1}{3} D_Y \frac{|F_\phi|^2}{(4\pi^4)} \ ,\nnn
&& \delta \tilde{m}_U^2 = -\frac{4}{3}D_Y \frac{|F_\phi|^2}{(4\pi^4)} \ ,\nnn
&& \delta \tilde{m}_D^2 = \frac{2}{3}D_Y \frac{|F_\phi|^2}{(4\pi^4)} \ ,\nnn
&& \delta \tilde{m}_{H_1}^2 = -D_Y \frac{|F_\phi|^2}{(4\pi^4)}  \ ,\nnn
&& \delta \tilde{m}_{H_2}^2 = D_Y \frac{|F_\phi|^2}{(4\pi^4)} \ , \label{eq:dtermcont}
\end{eqnarray}
where $D_Y$ comes from the common parameter $\xi$ in eq. (\ref{eq:dterm-xi}).

We can now evaluate the soft breaking terms of the nMSSM at the messenger scale using eq. (\ref{eq:formula}). 
We solve the renormalization group equations (RGE) using them as the boundary conditions, and then evaluate the soft breaking
terms at the weak scale. We use RGE codes contained in the NMSSMTools software package\cite{nmssmtools1}\cite{nmssmtools2}.
We also add the D-term contributions in eq. (\ref{eq:dterm-xi}) to the soft breaking mass.
In the next section, we discuss the phenomenology of the nMSSM with the soft breaking terms obtained by deflected anomaly mediation. 


\section{Phenomenology}
In this section, we investigate the phenomenological aspects of the nMSSM. 
First, we discuss the existence of Landau poles.
We demand that $\lambda$ should not meet the Landau pole up to the scale at
which the tadpoles are generated.
Next we study the regions of parameter space where successful electroweak breaking occurs, and we evaluate 
the mass of the lightest Higgs. Subsequently, we discuss the lightest 
neutralino as a dark matter candidate. We evaluate the relic density of the
lightest neutralino. We also discuss the direct detection of dark matter. Finally, we obtain sparticle mass spectra.

\subsection{The Landau pole}
In the nMSSM, tadpoles generated by supergravity interaction are proportional to powers of $\lambda$\cite{nmssm3}.
Therefore to maintain the tadpoles at the weak scale, $\lambda$ should not meet 
the Landau pole up to the scale, at which the tadpoles are generated.
We investigate the region of $\lambda$ and $\tan\beta$
 that satisfies the perturbativity condition below the GUT scale.

The beta-functions of $\lambda$ and $y_t$ are
\begin{eqnarray}
\beta_\lambda = \frac{1}{16\pi^2}\left(4\lambda^2 + 3 y_t^2 + 3 y_b^2 + y_\tau^2 -\frac{3}{5}g_1^2 -3 g_2^2 \right) \lambda \ ,  \nnn
\beta_{y_t} = \frac{1}{16\pi^2}\left(\lambda^2 + 6 y_t^2 + y_b^2 -\frac{13}{15}g_1^2 -3 g_2^2 -\frac{16}{3} g_3^2 \right) y_t \ .
\end{eqnarray}
These beta-functions strongly depend on $\lambda$ and $\tan \beta$ 
through the top Yukawa coupling.
Figure \ref{fig:lpole} shows the allowed region where the perturbativity 
is satisfied up to the GUT scale. 
The calculation is performed using the RGE code included in the NMSSMTools package. The current experimental value of the top mass is 
$173.1 \pm 1.3$ GeV\cite{topmass}. We take the central value for $m_{\rm top}$ as $173.1$ GeV. 
The shaded region is consistent with the perturbativity of $\lambda$. The result depends on 
the value of $m_{\rm top}$ and supersymmetric threshold corrections of $\alpha_s$.
Therefore there is small difference among the results of \cite{nmssm4} and \cite{nmssm_ewbg1} and our results.
In our calculation, the region where $\tan\beta \gtrsim 2.0$ and $\lambda \lesssim 0.7$ is allowed.

\FIGURE[htbp]{
\epsfig{file=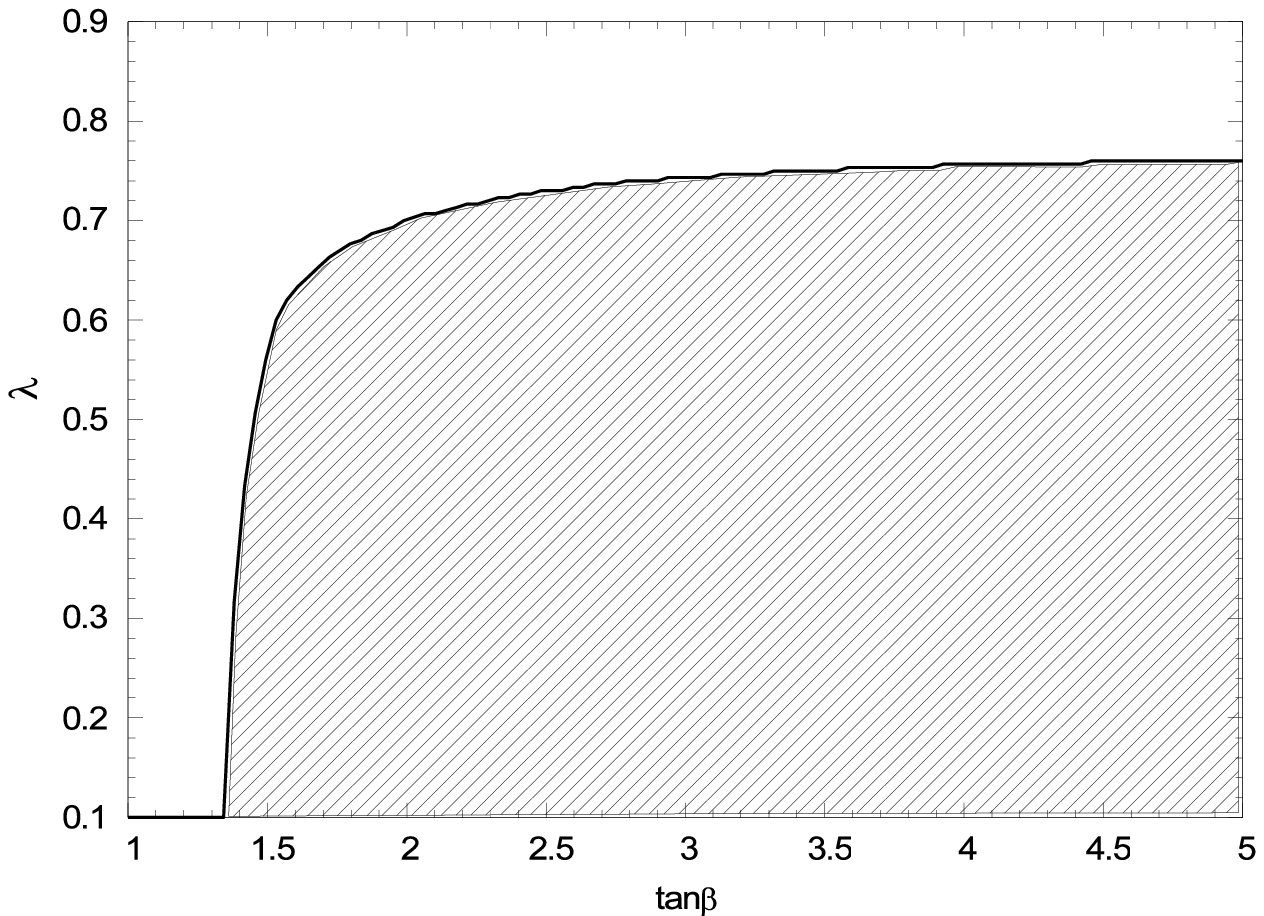, width=0.6\hsize}
\caption{
The region that is consistent with the perturbativity of $\lambda$ up to the GUT scale is shown.
The gray shaded region below the solid line is allowed. The region above the solid line is excluded owing to the existence of the Landau pole below the GUT scale.
}
\label{fig:lpole}
}

\subsection{Electroweak symmetry breaking}
In this subsection, we consider the conditions for electroweak 
symmetry breaking and evaluate $\mu_{eff} = \lambda \left<S\right>$. We 
also evaluate the mass of the lightest Higgs. 

After obtaining the soft breaking terms at the weak scale, we now evaluate the Higgs potential, $V = V_{tree} + \Delta V$.
From eqs. (\ref{eq:nmssm_sp}) and (\ref{eq:nmssm_soft}), the tree-level Higgs 
potential is written as 
\begin{eqnarray}
V_{tree} &=& \tilde{m}_{H_1}^2 H_1^\dagger H_1 + \tilde{m}_{H_2}^2 H_2^\dagger H_2 + m_s^2 |S|^2 + m_{12}^2 (H_1\cdot H_2 + h.c.) \nn \\
&+& \lambda^2 |H_1 \cdot H_2 |^2 + \lambda^2 |S|^2 (H_1^\dagger H_1 +H_2^\dagger H_2 ) + \frac{g^2}{2}|H_1^\dagger H_2|^2 \nn \\
&+& \frac{\bar{g}^2}{8}(H_2^\dagger H_2 - H_1^\dagger H_1)^2 + (t_s S + 
 h.c.) +(a_\lambda S H_1 \cdot H_2 + h.c.)  \ , \label{eq:potential}
\end{eqnarray}
where $\bar{g}^2 = g^2+g'^2$. 
$\Delta V$ is the one-loop contribution to the effective potential \cite{cw-potential}:
\begin{eqnarray}
\Delta V = \frac{1}{64\pi^2}\left(\sum_b g_b m_b^4 \left[\ln\left(\frac{m_b^2}{Q^2}\right)-\frac{3}{2}\right]
- \sum_f g_f m_f^4 \left[\ln\left(\frac{m_f^2}{Q^2}\right)-\frac{3}{2}\right]\right) \ . \label{eq:coleman-wein}
\end{eqnarray}
$g_b$ and $g_f$ are the degrees of freedom for bosons and fermions respectively.
We determine $\mu_{eff}\equiv \lambda \left<S\right>$, $t_s$ and 
$m_{12}^2$ using the stationary conditions of the Higgs potential. 
From eqs. (\ref{eq:potential}) and (\ref{eq:coleman-wein}), the stationary conditions are
\begin{eqnarray}
 \frac{\partial V}{\partial v_1} &=& 2v_1 \left[\tilde{m}_{H_1}^2 + (m_{12}^2 + 
					   a_\lambda 					   
v_s)\frac{v_2}{v_1}-\frac{\bar{g}^2}{4}(v_2^2-v_1^2)+\lambda^2(v_2^2+v_s
^2) + \frac{1}{2v_1}\frac{\p \Delta V}{\p v_1} \right] = 0 \ , \nn \\
 \frac{\partial V}{\partial v_2} &=& 2v_2 \left[\tilde{m}_{H_2}^2 + (m_{12}^2 + 
					   a_\lambda v_s)\frac{v_1}{v_2}+\frac{\bar{g}^2}{4}(v_2^2-v_1^2)+\lambda^2(v_1^2+v_s
^2) + \frac{1}{2v_2}\frac{\p \Delta V}{\p v_2} \right] =0  \ ,\nn \\
 \frac{\partial V}{\partial v_s} &=& 2v_s \left[m_s^2+\lambda^2(v_1^2+v_2^2)+\frac{t_s}{v_s}+a_\lambda \frac{v_1 v_2}{v_s} + \frac{1}{2v_s}\frac{\p \Delta V}{\p v_s} \right]=0 \label{eq:higgs_kyokuchi} \ ,
\end{eqnarray}
where $v_1=\left<H_1^0\right>$, $v_2=\left<H_2^0\right>$ and $v_s = \left<S\right>$. 
As we describe later, there is only a small region
of parameter space where successful electroweak symmetry breaking 
occurs with $v_s > 0$; therefore, we take $v_s < 0$.
From eq. (\ref{eq:higgs_kyokuchi}), $\mu_{eff}$ can be determined by,
\begin{eqnarray}
\mu_{eff}^2 = -\frac{M_Z^2}{2} + \frac{\tilde{m}_{H_1}^2 + \frac{1}{2v_1}\frac{\partial \Delta V}{\partial v_1} - 
\left(\tilde{m}_{H_2}^2 + \frac{1}{2v_2}\frac{\partial \Delta V}{\partial v_2}\right) \tan^2\beta}{\tan^2\beta-1} \label{eq:mueff}.
\end{eqnarray}
$\mu_{eff}$, $m_{12}^2$ and $t_s$ are determined from eq. (\ref{eq:higgs_kyokuchi}). We now evaluate the Higgs mass. We 
expand $H_1^0$, $H_2^0$ and $S$ as
\begin{eqnarray}
H_1^0 &=& v_1 + \frac{1}{\sqrt{2}}\left(h_1^0 + i a_1\right), \nnn
H_2^0 &=& v_2 + \frac{1}{\sqrt{2}}\left(h_2^0 + i a_2\right), \nnn
S^0 &=& v_s + \frac{1}{\sqrt{2}}\left(s + i a_s\right). 
\end{eqnarray}
Using these expanded fields, the CP-even Higgs mass matrix is written as
\begin{eqnarray}
\left(h_1^0 \ h_2^0 \ S \right) M^2 \left(
\begin{array}{c}
h_1^0 \\
h_2^0 \\
S
\end{array}
\right) .
\end{eqnarray}
At tree level, the components of $M^2$ are
\begin{eqnarray}
M_{11}^2 &=& s_\beta^2 M_a^2 + c_\beta^2 M_Z^2 \ , \nnn
M_{12}^2 &=& -s_\beta c_\beta \left(M_a^2 + M_Z^2 - 2\lambda^2 v^2 \right) \ ,\nnn
M_{13}^2 &=& v \left(s_\beta a_\lambda + 2 c_\beta \lambda^2 v_s^2 \right) \ ,\nnn
M_{22}^2 &=& c_\beta^2 M_a^2 + s_\beta^2 M_Z^2 \ ,\nnn
M_{23}^2 &=& v \left(c_\beta a_\lambda 2 + s_\beta \lambda^2 v_s \right) \ ,\nnn
M_{33}^2 &=& -\frac{1}{v_s}\left(t_s + s_\beta c_\beta a_\lambda v_s \right),
\end{eqnarray}
where $c_\beta = \cos\beta$ and $s_\beta = \sin\beta$.
The CP-odd Higgs mass matrix at tree-level is
\begin{eqnarray}
\left(A^0 \ a_s \right)
\left[
\begin{array}{cc}
M_a^2 & -a_\lambda v_s \\
-a_\lambda v_s & -\frac{1}{v_s}\left( t_s + s_\beta c_\beta a_\lambda v^2 \right)
\end{array}
\right]
\left(
\begin{array}{c}
A^0 \\
a_s
\end{array}
\right) ,
\end{eqnarray}
where $M_a^2 = -\left(m_{12}^2 + a_\lambda v_s \right)/c_\beta s_\beta$. 
$A^0 = a_d s_\beta + a_u c_\beta$, and its orthogonal combination is absorbed by the Z boson.

We now present the results of numerical calculations.
Figure \ref{fig:nmssm_ewsbok} shows the allowed region of successful electroweak symmetry breaking without tachyonic sleptons.
We set the messenger scale to $5 F_\phi \simeq 150 \ {\rm TeV}$.
Successful electroweak symmetry breaking occurs in the region covered by red squares.
In the region covered by blue crosses, the mass of the lightest Higgs 
satisfies the LEP bound with the electroweak symmetry breaking.
When the number of the messengers, $N_f$ increases, the allowed region of the 
deflection parameter $d$ is shifted downward. Therefore, in the scenario with small 
$d$ ($d < 1$), two or more messengers have to exist. 
For simplicity, we assume that there is one messenger in the following analysis. 
Although there is a region where successful electroweak symmetry 
breaking occurs with large $\tan\beta$ and $v_s >0$, 
the region is very small. Therefore we take $v_s < 0$.

Figure \ref{fig:nmssm_mueff} shows the dependence of $\mu_{eff}$ on SUSY 
breaking. 
We see that $|\mu_{eff}|$ is a decreasing function of $D_Y$, while it is an increasing function of $d$.
This can be understood from eqs. (\ref{eq:formula}), (\ref{eq:dtermcont}) and (\ref{eq:mueff}).
When $D_Y$ increases, 
$m_{H_1}^2$ decreases and $m_{H_2}^2$ increases. This implies that $|\mu_{eff}|$ decreases as $D_Y$ increases.
When $d$ increases, 
$m_{H_1}^2$ and $m_{H_2}^2$ increase at almost the same rate. This implies that $|\mu_{eff}|$ increases as $d$ increases.
In this scenario, moderate values of $\mu_{eff}$, $100 < |\mu_{eff}| < 
550$, are obtained without meeting the Landau pole.

Figure \ref{fig:higgsmass} shows the dependence of the lightest Higgs mass on $d$ and $D_Y$. 
The calculation is performed with NMSSMTools, including two-loop corrections. We extend the codes to include tadpoles. 
In this scenario, the mass of the lightest Higgs can be
heavier than the LEP bound.

\FIGURE[htbp]{
\hspace*{-6mm}
\epsfig{file=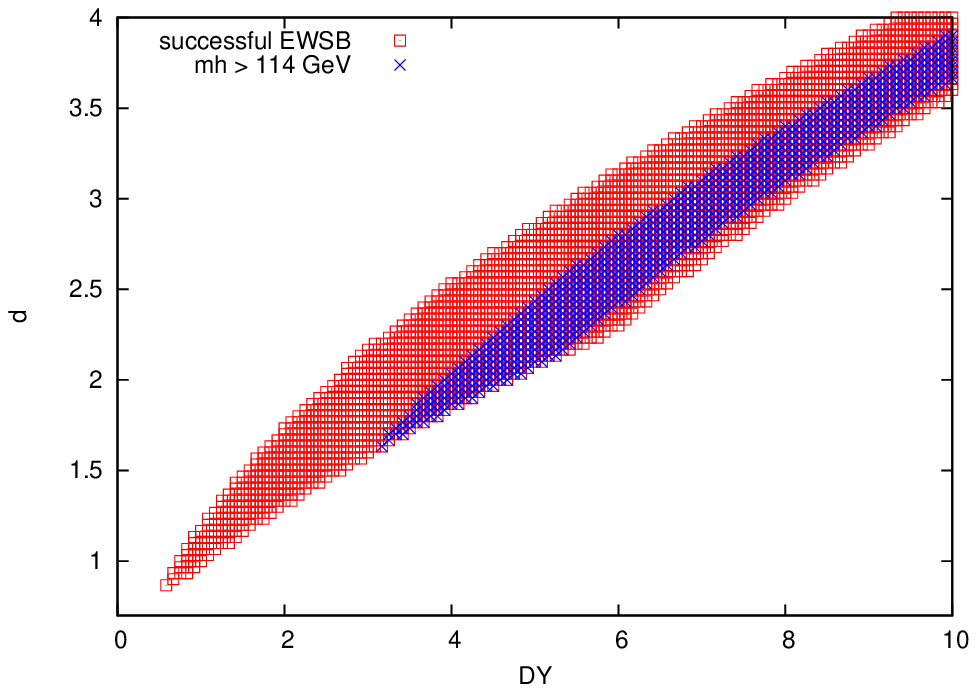,width=0.45\hsize}
\epsfig{file=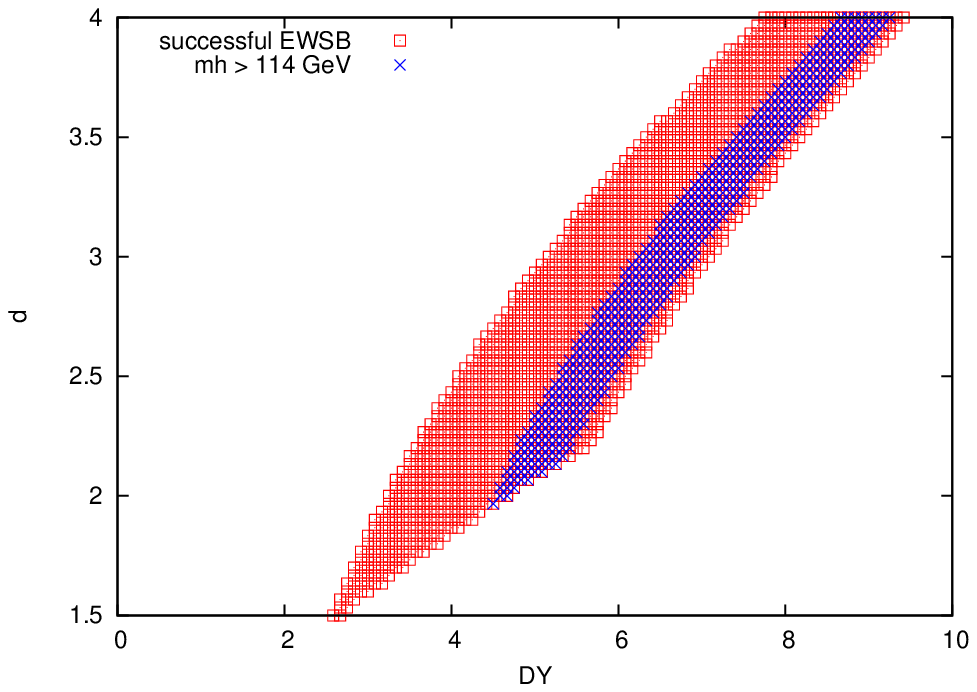,width=0.45\hsize}
\epsfig{file=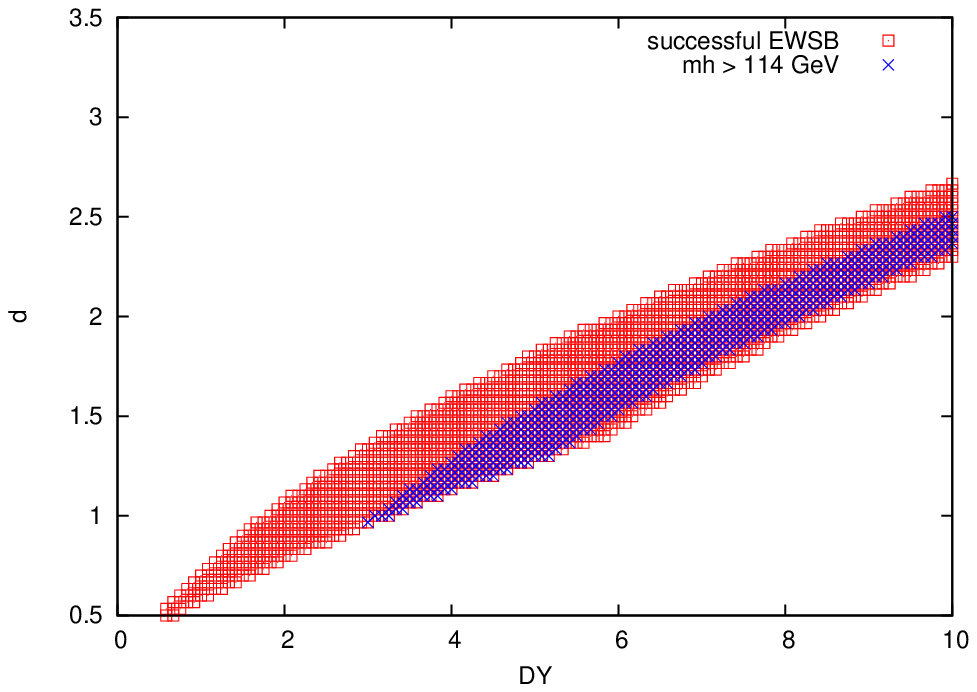,width=0.45\hsize}
\epsfig{file=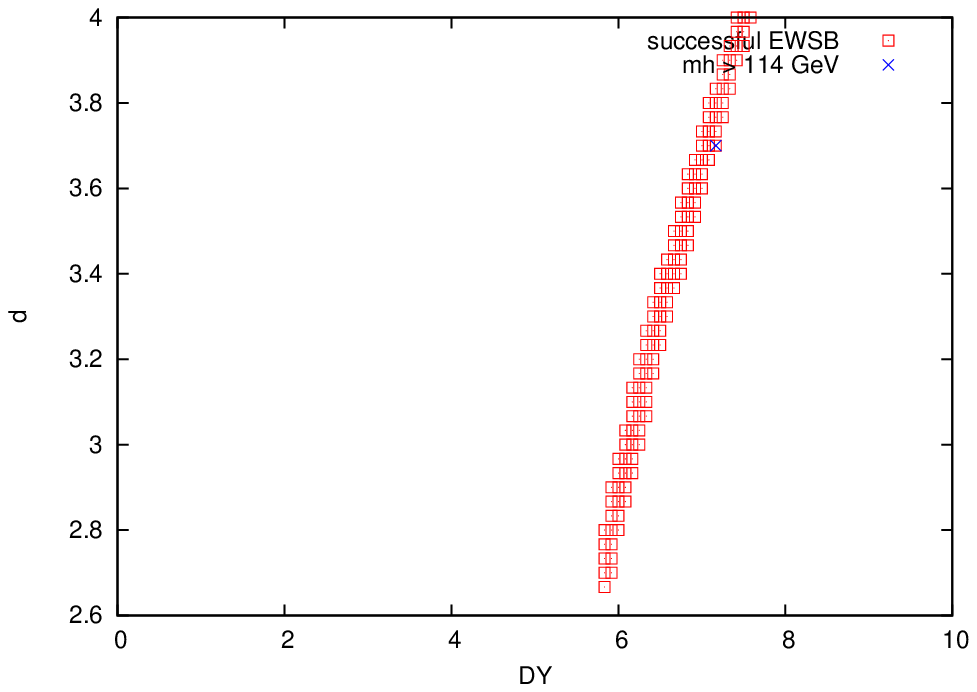,width=0.45\hsize}
\caption{
Successful electroweak symmetry breaking occurs in the region covered by red squares, and the region covered by blue crosses satisfies
the Higgs mass bound of the LEP. In other regions, the sleptons are tachyonic.
The calculation is performed with $\lambda=0.69$ and $m_0 = F_\phi/(4\pi)^4 = 200 \ {\rm GeV}$. The messenger scale is taken to be $5 F_\phi$.
$\tan\beta$ and the number of messengers $N_f$ are $\tan\beta=2$ and $N_f=1$ in the top-left figure, 
$\tan\beta=3$ and $N_f=1$ in the top-right figure,
$\tan\beta=2$ and $N_f=2$ in the bottom-left figure and
$\tan\beta=20$ and $N_f=1$ in the bottom-right figure. The bottom-right figure is evaluated with $v_s > 0$. 
The others are evaluated with $v_s < 0$.
}
\label{fig:nmssm_ewsbok}
}

\FIGURE[htbp]{
\epsfig{file=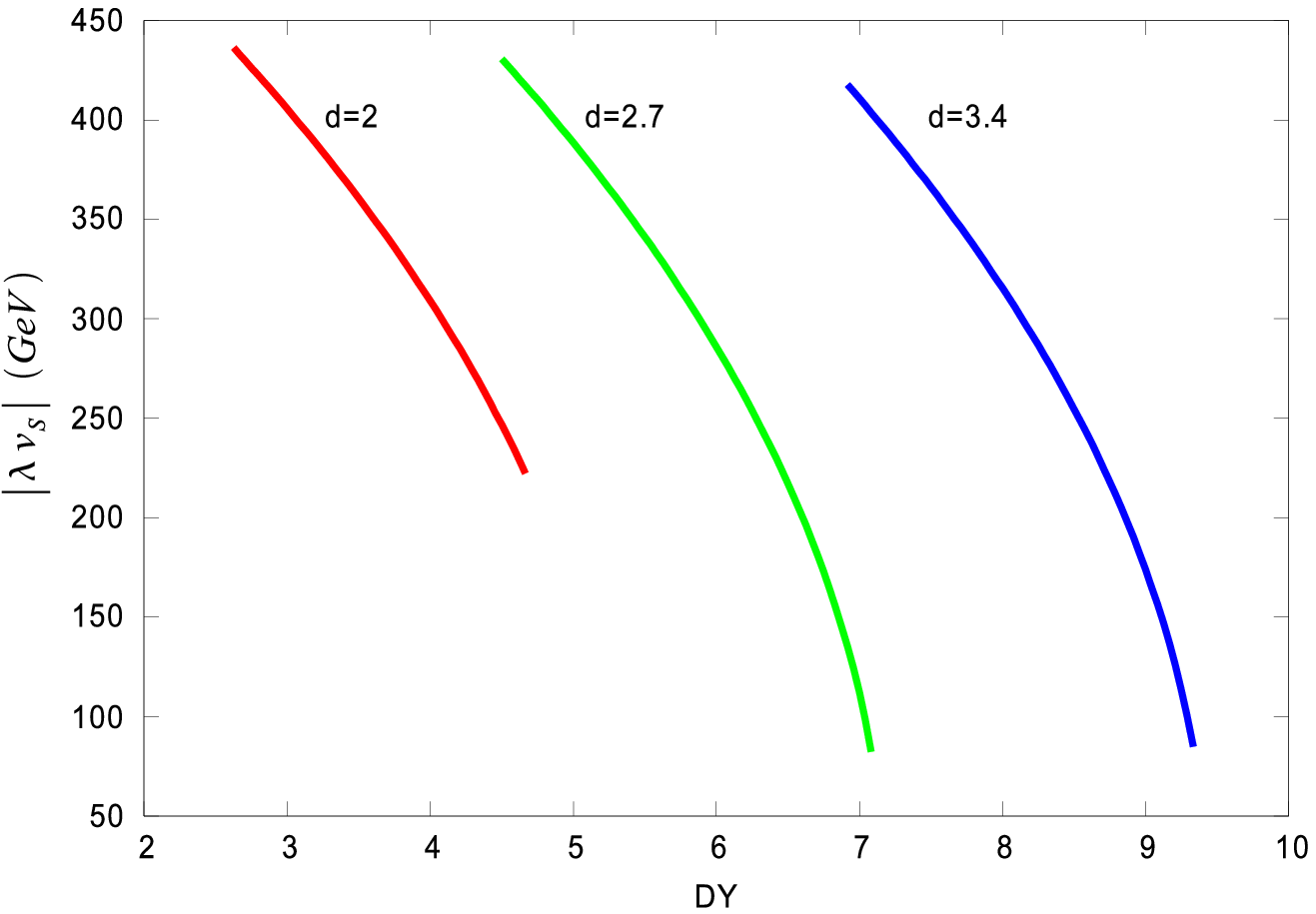,width=0.45\hsize}
\epsfig{file=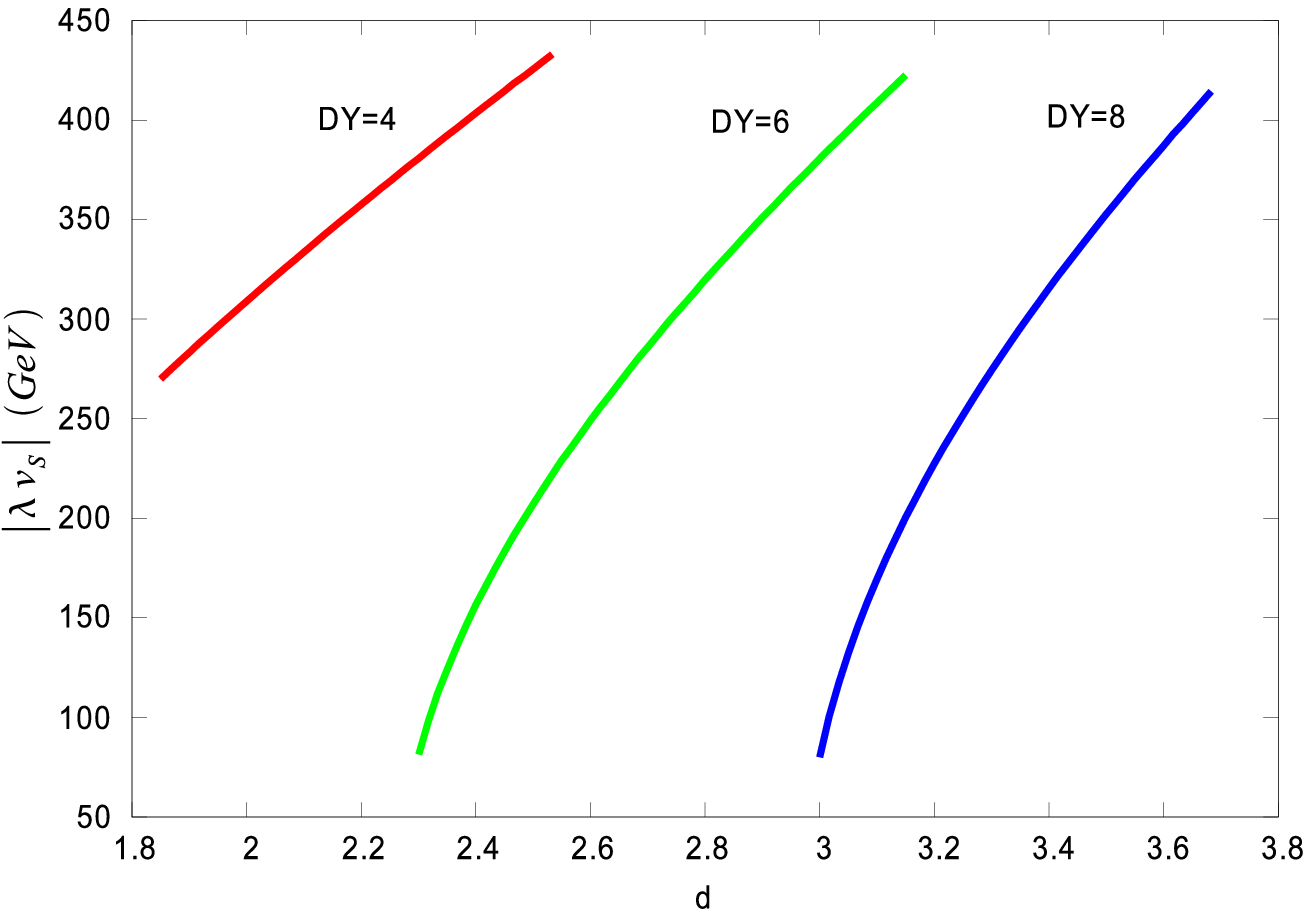,width=0.45\hsize}
\caption{
The values of $|\lambda v_s|$ are shown. The calculations are performed with $\lambda=0.69$, $\tan\beta=2$ and $m_0=200 \ {\rm GeV}$.
Moderate values of $|\lambda v_s|$ are obtained.
}
\label{fig:nmssm_mueff}
}
 
\FIGURE[htbp]{
 \epsfig{file=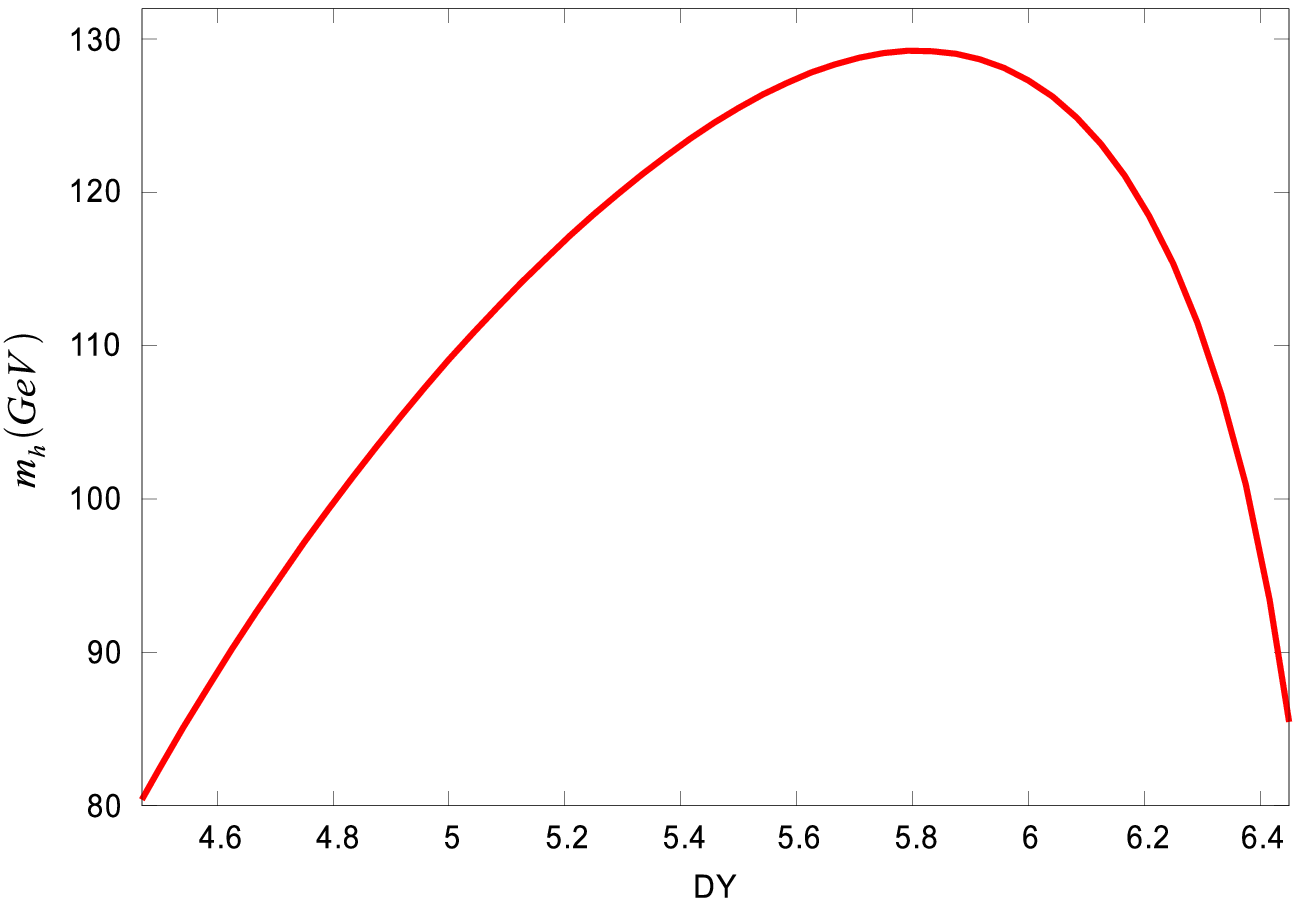,width=0.45\hsize}
 \epsfig{file=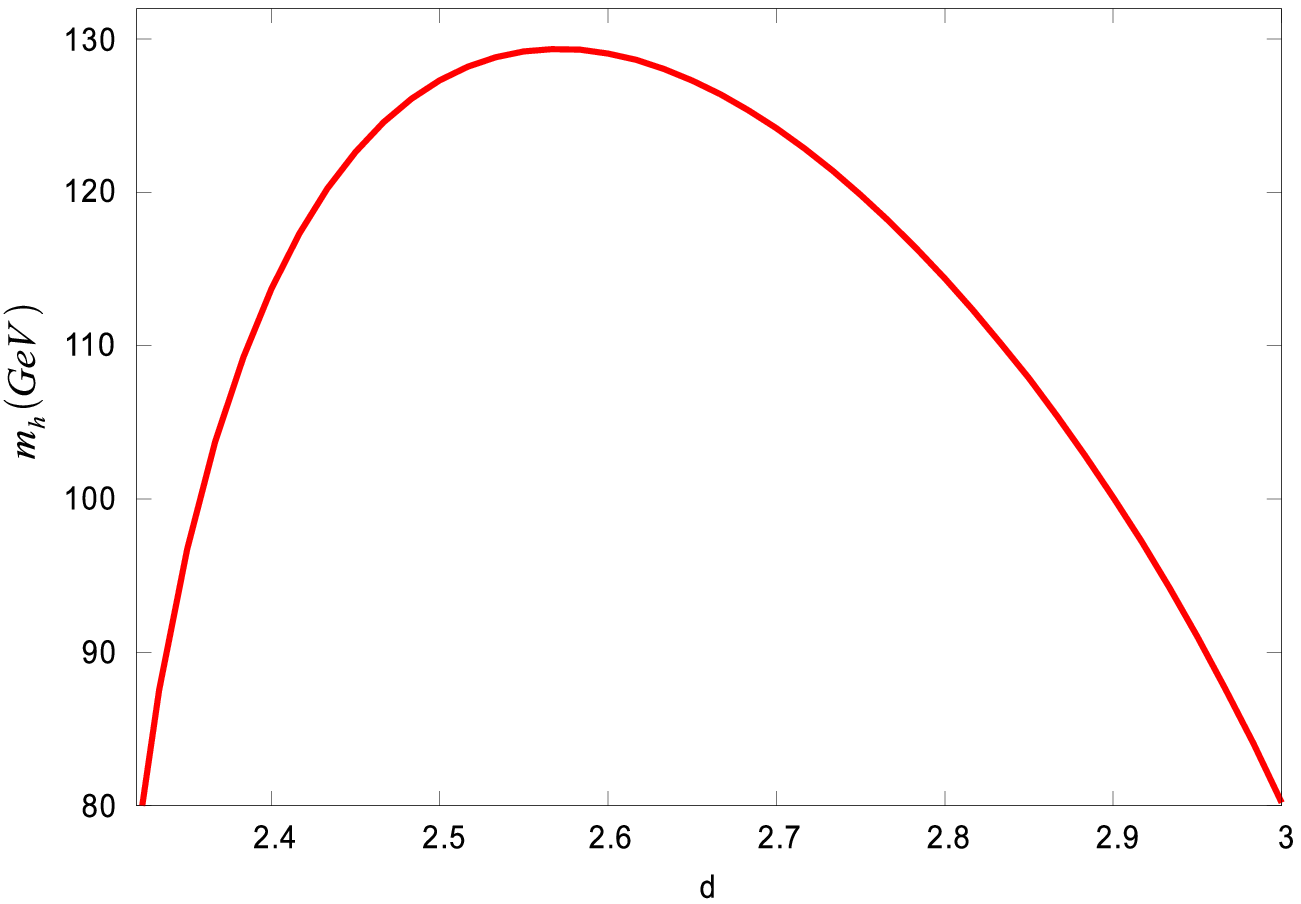,width=0.45\hsize}
\caption{
The dependence of the Higgs mass on the SUSY breaking parameter is shown. In the left figure we set $d=2.5$, and in the right figure we set $D_Y = 6$. 
Other parameters are chosen as $\lambda=0.69, \tan\beta=2.0$ and $m_0=200 {\rm GeV}$ in both figures.
}
\label{fig:higgsmass}
}

\subsection{Dark matter}
In this scenario, the lightest neutralino is the LSP in the wide range of parameter space. 
Therefore, the lightest neutralino is a candidate for dark matter. In this 
subsection, we evaluate the relic density of the lightest neutralino, 
which is mainly composed of a singlino. We also calculate the the neutralino-proton 
scattering cross section, and discuss the direct detection of dark matter.

In the nMSSM, the relic density of the lightest neutralino strongly depends 
on its mass\cite{nmssm1}\cite{nmssm2}. Although the dominant contribution to the
annihilation cross section is s-channel $Z$ boson exchange, the coupling 
between the $Z$ boson and $\tilde{N}_1$ is significantly small. This is 
because the lightest neutralino, $\tilde{N}_1$ is mainly composed of the 
fermionic component of the nMSSM gauge singlet, $\hat{S}$. 
The resonant effect near the $Z$ pole mass is important for the sufficient
annihilation of the lightest neutralino.

The neutralino mass matrix is
\begin{eqnarray}
\left(\tilde{B} \ \tilde{W} \ \tilde{H}_1^0 \ \tilde{H}_2^0 \ \tilde{S}\right)
\left(
\begin{array}{ccccc}
m_{\lambda_1} & 0 & -c_\beta s_w M_Z & s_\beta s_w M_Z & 0 \\
0 & m_{\lambda_2} & c_\beta c_w M_Z & -s_\beta c_w M_Z & 0 \\
-c_\beta c_w M_Z & c_\beta c_w M_Z & 0 & \mu_{eff} & \lambda v_2 \\
s_\beta s_w M_Z & -s_\beta c_w M_Z & \mu_{eff} & 0 & \lambda v_1 \\
0 & 0 & \lambda v_2 & \lambda v_1 & 0 
\end{array}
\right)
\left(
\begin{array}{c}
\tilde{B} \\
\tilde{W} \\
\tilde{H}_1^0 \\
\tilde{H}_2^0 \\
\tilde{S}
\end{array}
\right)
,
\end{eqnarray}
where $s_\beta = \sin\beta$, $c_\beta = \cos\beta$ and $s_w = \sin{\theta_W}$.
$\tilde{B}$, $\tilde{W}$, $\tilde{H}_{1,2}^0$ and 
$\tilde{S}$ denote the bino, wino, higgsino and singlino respectively. 
The mass of the lightest neutralino, $m_{\chi_1}$, becomes heavier
as $|\mu_{eff}|$ decrease. This is because the mixing of the higgsinos becomes small as one can see from eq. (4.10).
Since $|\mu_{eff}|$ is a decreasing function of $D_Y$, a larger $D_Y$ leads to a larger $m_{\chi_1}$. 
The dependence of the lightest neutralino mass on $D_Y$ is shown in Fig.\ref{fig:relic}.
On the other hand,
a larger $d$ leads to a smaller $m_{\chi_1}$.

Figure \ref{fig:relic} shows $m_{\chi}$ and the relic density of the neutralino, $\Omega_{\chi} h^2$. $m_{\chi}$ and $\Omega_{\chi} h^2$ are calculated 
with NMSSMTools and micrOMEGAs\cite{microomegas}\cite{microomegas2}. When $m_{\chi}$ is large and close to $m_Z$, 
$\Omega_{\chi} h^2$ is small. The observed relic density of dark matter is given by\cite{wmap1}\cite{wmap2}
\begin{eqnarray}
0.094 < \Omega_{CDM} h^2 < 0.136 .
\end{eqnarray}
This condition is satisfied with $m_{\chi} \simeq 35$ GeV. With such light dark matter, there are strong limits for the spin-independent
 WIMP-nucleon scattering cross section from CDMS\cite{cdms} and XENON10\cite{xenon}. The strongest limit for
 the cross section is for it to be less than $5\times 10^{-44} {\rm cm}^2$ for $m_{\chi_1} \simeq 30 \ {\rm GeV}$.

\FIGURE[htbp]{
 \epsfig{file=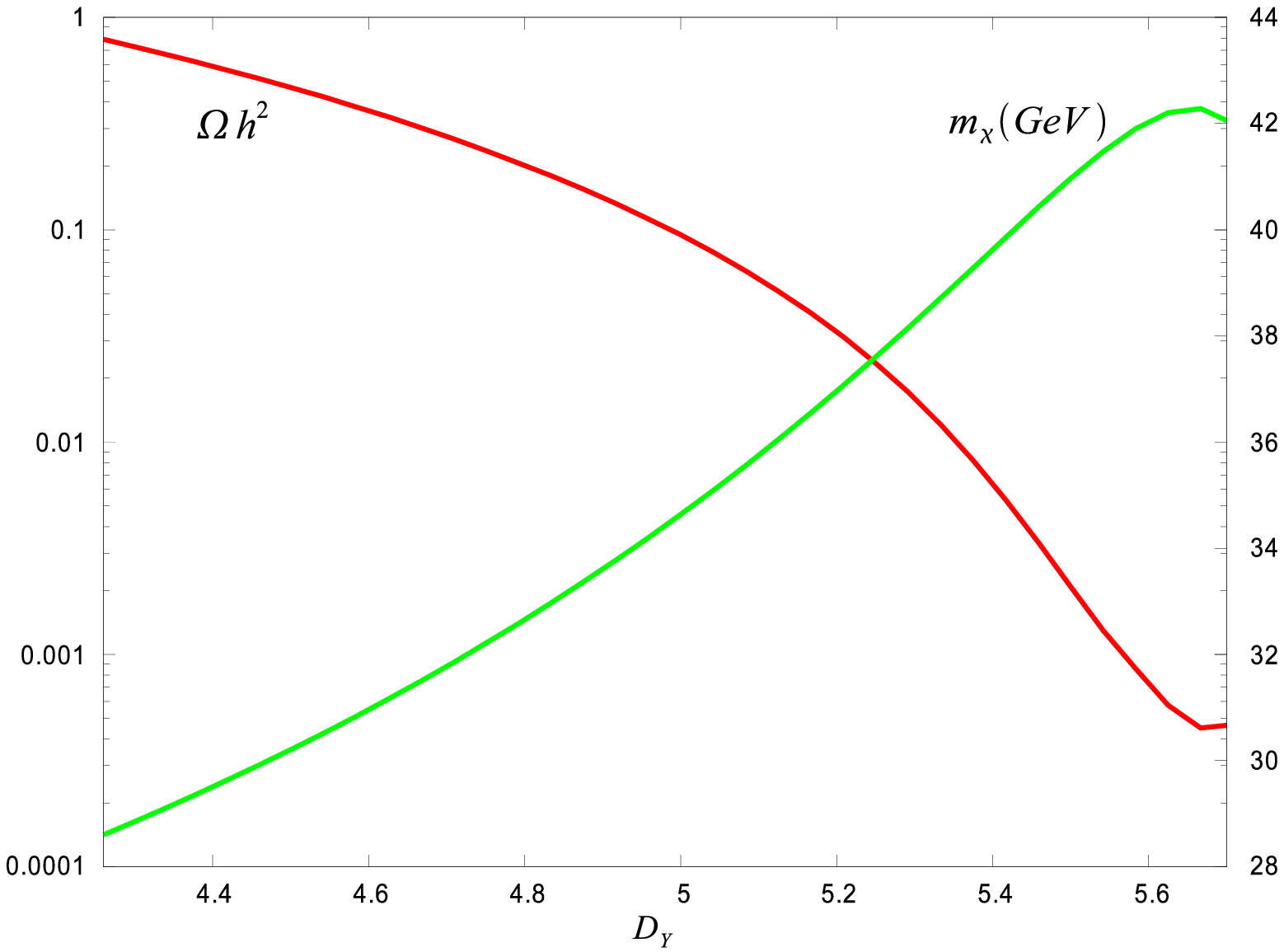,width=0.6\hsize}
\caption{
The mass and the relic density of the lightest neutralino are shown as functions of the 
SUSY breaking parameter $D_Y$. The other parameters are chosen as $m_0 = 200$ GeV, $d=2.25, \lambda=0.69$ and $\tan\beta=2.0$
}
\label{fig:relic}
}

The spin-independent WIMP-nucleon elastic scattering cross section is 
written as
\begin{eqnarray}
\sigma^{\rm SI} = \frac{4 m_{\chi}^2 m_{nucleus}^2}{\pi (m_{\chi} + 
 m_{nucleus})^2} \left[Z f_p +(A-Z)f_n\right]^2\ .
\end{eqnarray}
$f_{p,n}$ is the coupling between the WIMP and a nucleon given by \cite{susy_darkmatter}
\begin{eqnarray}
f_{p,n} = \sum_{q=u,d,s} f_{T_q}^{(p,n)} a_q \frac{m_{p,n}}{m_q} + \frac{2}{27} f_{T_G}^{(p,n)} \sum_{q=c,b,t} a_q \frac{m_{p,n}}{m_{q}} .
\end{eqnarray}
$a_q$ are the WIMP-quark couplings. We focus on the dark matter-proton scattering cross section in the following discussion. 
The parameter $f_{T_q}$ is defined by 
\begin{eqnarray}
m_p f_{T_q} \equiv m_q \left<N|\bar{q}q|N\right> \equiv m_q B_q ,
\end{eqnarray}
and $f_{TG}=1-\sum_{q=u,d,s} f_{T_q}$. $f_{T_q}$ can be written as 
\cite{direct_dm_update}
\begin{eqnarray}
f_{T_u} &=& \frac{m_u B_u}{m_p} = \frac{2\sigma_{\pi N}}{m_p \left(1+ \frac{m_d}{m_u}\right)\left(1+  \frac{B_d}{B_u}\right)} \ ,\nnn
f_{T_d} &=& \frac{m_d B_d}{m_p} = \frac{2\sigma_{\pi N}}{m_p \left(1+  
    \frac{m_u}{m_d}\right)\left(1+  \frac{B_u}{B_d}\right)} \ ,\nnn
f_{T_s} &=& \frac{m_s B_s}{m_p} = \frac{ y \left(\frac{m_s}{m_d}\right)  \sigma_{\pi N}  }{m_p \left(1+  \frac{m_u}{m_d}\right)} , \label{eq:ftq}
\end{eqnarray}
where $\sigma_{\pi N}$ is the $\pi$-nucleon sigma term:
\begin{eqnarray}
\sigma_{\pi N} = \frac{1}{2}\left(m_u + m_d\right)\left(B_u + B_d\right) .
\end{eqnarray}
The phenomenological value of $\sigma_{\pi N}$ is $64\pm 8$ MeV\cite{direct_dm_update}.
$y$ denotes the ratio of the strange quark component in the nucleon, defined as
\begin{eqnarray}
y = \frac{2 B_s}{B_u + B_d}\ .
\end{eqnarray}
$y$ can be determined by the relation,
\begin{eqnarray}
\sigma_{0} = \sigma_{\pi N} \left(1-y\right) = \frac{1}{2}\left(m_u + m_d \right)\left(B_u + B_d -2B_s \right) .
\end{eqnarray}
$\sigma_0$ can be evaluated from baryon mass spectra using chiral perturbation theory. From \cite{sigma_0}, $\sigma_0 = 36 \pm 7$ MeV.
There is large ambiguity for $y$. When $(\sigma_{\pi N}, \sigma_0) = (64, 36)$ MeV, $y=0.44$. 
On the other hand, according to a recent lattice calculation \cite{ohki},
 $y$ has a small value such as $0.03$.

The ratios of the quark mass are taken from \cite{quark_mass_ratio}.
\begin{eqnarray}
\frac{m_u}{m_d} = 0.553 \pm 0.043, \  \frac{m_d}{m_s} = 18.9 \pm 0.8 . \label{eq:mumd}
\end{eqnarray}
The ratios of the form factors are written as
\begin{eqnarray}
\frac{B_d}{B_u} = \frac{2 + (z-1)y}{2z-(z-1)y} , \label{bdbu}
\end{eqnarray}
where
\begin{eqnarray}
z = \frac{B_u-B_s}{B_d-B_s} .
\end{eqnarray}
$z$ can be calculated from the baryon mass, and its value is 
1.49\cite{dm_zparameter}. We can now determine $f_{T_q}$ from eqs. 
(\ref{eq:ftq}), (\ref{eq:mumd}) and (\ref{bdbu}). 
When $y=0.44$,
\begin{eqnarray}
f_{T_u} \approx 0.027, \ \ f_{T_d} \approx 0.039, \ \ f_{T_s} \approx 0.365, \ \ f_{TG} \approx 0.569,
\end{eqnarray}
and when $y=0.03$,
\begin{eqnarray}
f_{T_u} \approx 0.029, \ \ f_{T_d} \approx 0.036, \ \ f_{T_s} \approx 0.025, \ \ f_{TG} \approx 0.91.
\end{eqnarray}
In these two cases, $f_{T_s}$ and $f_{TG}$ are very different. This 
affects the spin-independent cross section of the WIMP-nucleon scattering significantly.

WIMP-quark couplings, $a_q$, consist of two parts. One part arises from squark 
s-channel exchange and the other arises from the t-channel exchange of 
the neutral Higgs. 
The couplings from squark exchange are given by \cite{aq_squark}
\begin{eqnarray}
a_{q_i}^{\tilde{q}} = -\frac{1}{2(\tilde{m}_{1i}^2-m_{\chi}^2)} {\rm Re} \left[X_i Y_i^* \right] 
- \frac{1}{2(\tilde{m}_{2i}^2-m_{\chi}^2)} {\rm Re} \left[W_i 
						     V_i^*\right] \ ,
\end{eqnarray}
where
\begin{eqnarray}
X_i &=& \eta_{11}^* \frac{g m_{q_i} N_{1,5-i}^*}{2M_w B_i} - \eta_{12}^* 
 e_i g' N_{11}^*  \ ,\nnn
Y_i &=& \eta_{11}^* \left(\frac{y_i}{2} g' N_{11} + g T_{3i} N_{12}\right) + \eta_{12}^* \frac{g m_{q_i} N_{1,5-i}}{2M_w B_i} \ ,\nnn
W_i &=& \eta_{21}^* \frac{g m_{q_i} N_{1,5-i}^*}{2M_w B_i} - \eta_{22}^* e_i g' N_{11}^* \ ,\nnn
Y_i &=& \eta_{21}^* \left(\frac{y_i}{2} g' N_{11} + g T_{3i} N_{12}\right) + \eta_{22}^* \frac{g m_{q_i} N_{1,5-i}}{2M_w B_i},
\end{eqnarray}
and $i=1$ for an up-type quark and $i=2$ for a down-type quark. $\tilde{m}_{1i}$ and 
$\tilde{m}_{2i}$ denote a light squark mass and a heavy squark mass respectively. 
$\eta$ denotes a squark
mixing such that
\begin{eqnarray}
\tilde{q}_l = \eta_{l1} \tilde{q}_L + \eta_{l2} \tilde{q}_R .
\end{eqnarray}
$y_i$, $T_{3i}$ and $e_i$ denote the hypercharge, isospin and electric 
charge of the quarks respectively. $B_1 = \sin\beta$ and $B_2 = \cos\beta$.

The couplings from neutral Higgs exchange in the nMSSM are given by\cite{nmssm_aq}
\begin{eqnarray}
a_{q_i}^h = \sum_{a=1}^3 \frac{1}{m_{h_a^0}^2} {C_Y}_a^i {\rm Re} 
 [C_{H}^a] \ ,
\end{eqnarray}
where
\begin{eqnarray}
{C_Y}_a^i &=& -\frac{g m_{q_i}}{4 M_w B_i} S_{a,3-i} \ ,\nnn
{C_{H}^a} &=& \left(-g N_{12}^* + g' N_{11}^* \right)\left(S_{a1}N_{13}^* - S_{a2}N_{14}^* \right) \nnn
&& - \sqrt{2} \lambda \left[S_{a3}N_{13}^* N_{14}^* + N_{15}^* 
		       \left(S_{a2}N_{13}^* + S_{a1}N_{14}^* 
		      \right)\right] \ .\label{eq:a_from_higgs} 
\end{eqnarray}
$S_{ij}$ denotes Higgs mixing. One can write the mass eigenstate of the Higgs as
\begin{eqnarray}
h_a^0 = S_{a1} h_d^0 + S_{a2} h_u^0 + S_{a3} h_s .
\end{eqnarray}
When $\lambda=0$, eq. (\ref{eq:a_from_higgs}) agrees with the couplings 
in the MSSM given in \cite{aq_squark}.

Figure \ref{fig:si_cross} shows the spin-independent cross section as a function of $d$ and $D_Y$. When $y=0.44$, $\sigma^{SI}$ is already excluded
by the current experiments. On the other hand, when $y=0.03$, 
$\sigma^{SI}$ is smaller than the upper limit from XENON10 in many regions of the parameter space.
In this case, $\sigma^{SI}$ is large enough to be detected or be 
excluded by the next-generation experiments.

\FIGURE[htbp]{
 \epsfig{file=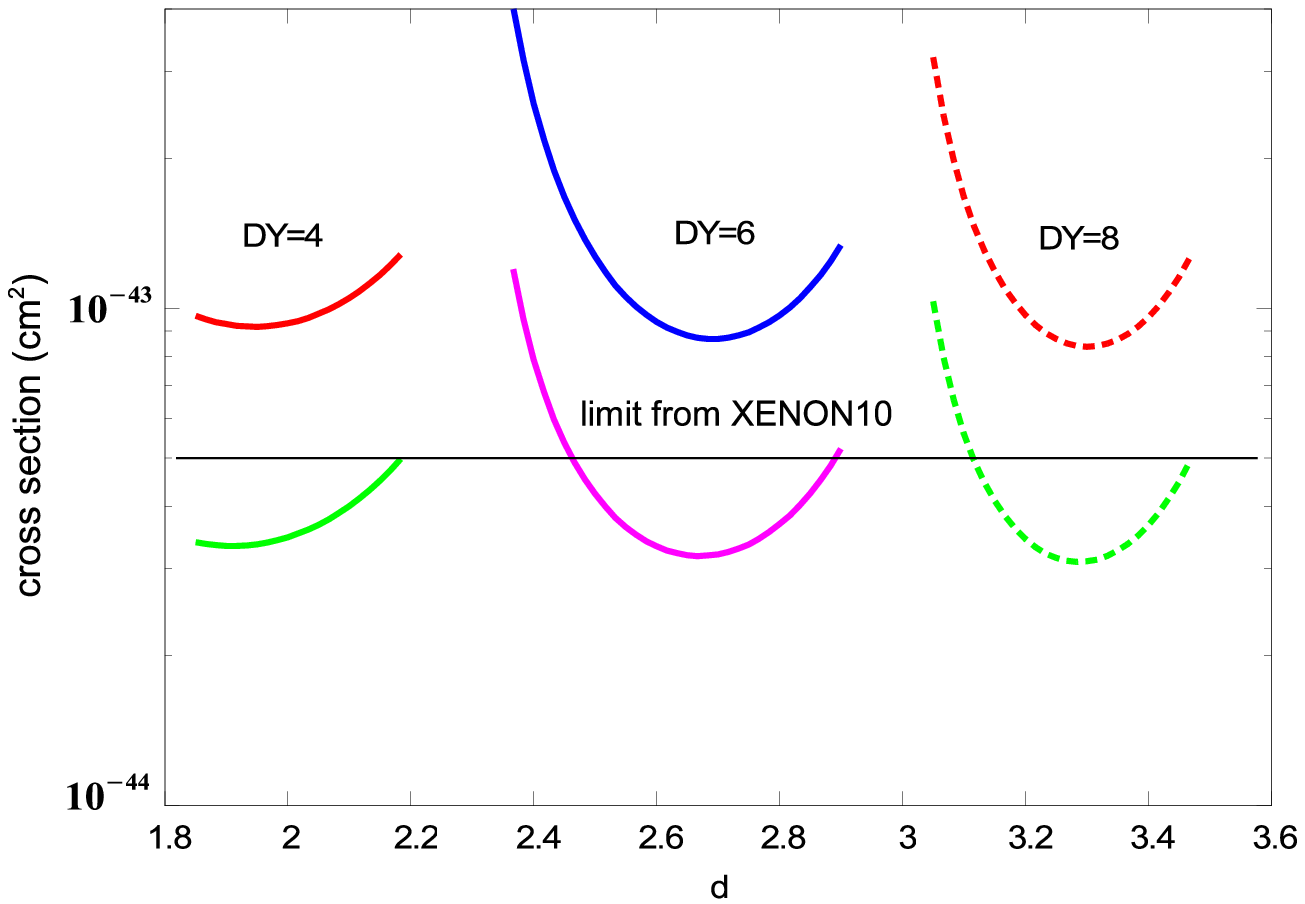,width=0.49\hsize}
 \epsfig{file=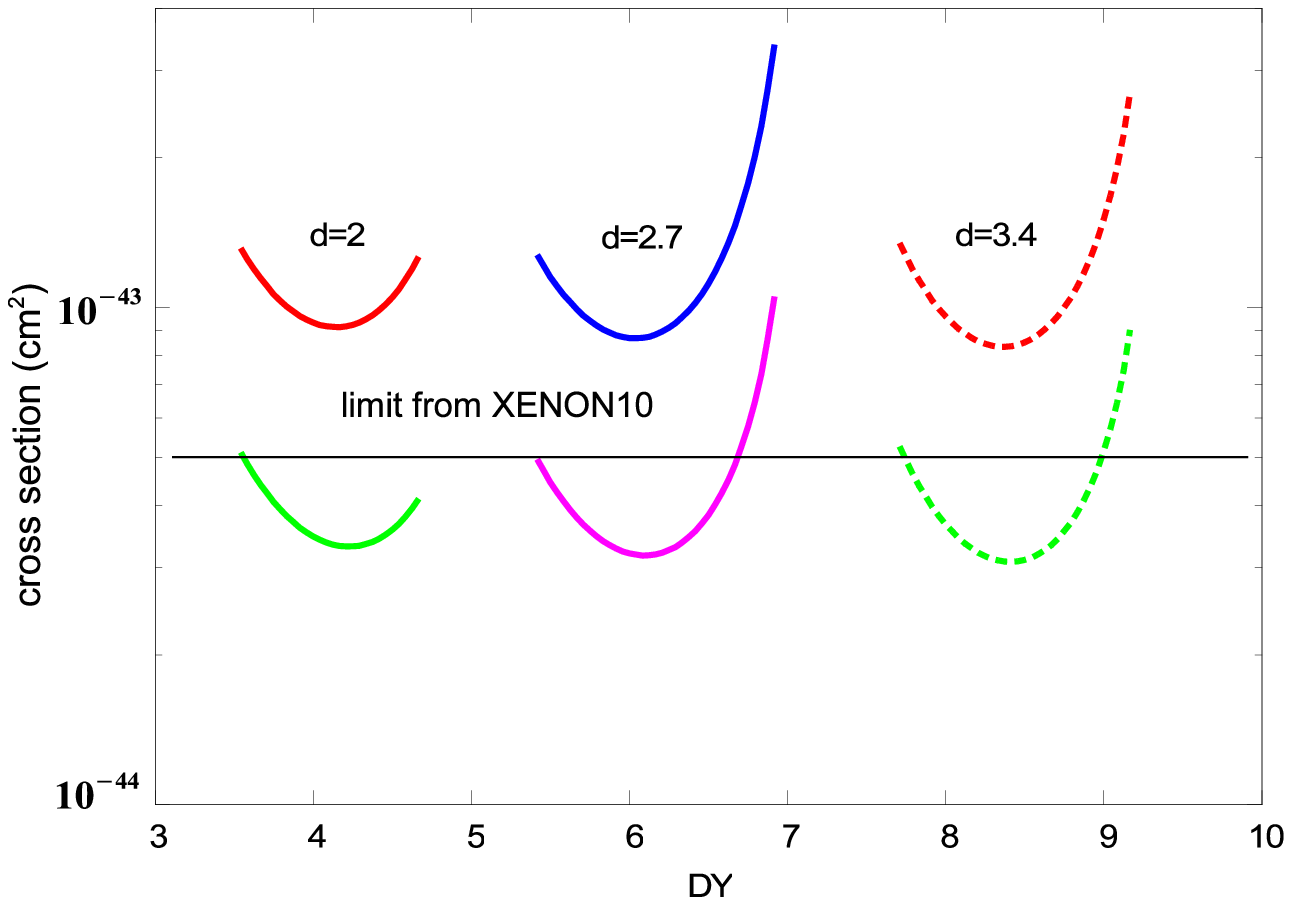,width=0.49\hsize}
\caption{
The spin-independent cross sections, $\sigma^{SI}$ are shown. $\sigma^{SI}$ are calculated with $m_0 = 200$ GeV, $\lambda=0.69$ and $\tan\beta=2.0$.
The upper three lines are calculated with $y=0.44$. $y$ is evaluated with chiral perturbation.
The lower three lines are calculated with $y=0.03$, which is the result from a recent lattice calculation.
}
\label{fig:si_cross}
}

\subsection{Mass spectrum}
Here we present the sparticle mass spectrum, the relic density of the 
neutralino and the spin-independent cross section of dark matter-proton scattering.
The mass spectra are calculated using NMSSMTools, and the relic 
densities are calculated using micrOMEGAs.

In this scenario, the gluino is light in a wide range of the parameter space. This is because the contributions from gauge mediation and anomaly mediation
 cancel. From eq. (\ref{eq:formula}), the gluino mass at the messenger scale 
 is written as
\begin{eqnarray}
m_{\ti{g}} = -g_3^2 \left(3-d\right)m_0 \label{eq:mass_gl} \ ,
\end{eqnarray}
where $m_0 = \displaystyle F_\phi/(16\pi^2)$. Particularly in the region where $d=3$, the gluino mass $m_{\ti{g}}$ vanishes.

The results of the numerical calculation are presented in Table \ref{tab:masses}. When the deflection parameter $d$ changes, 
the overall scale of the soft breaking terms changes. However, once we impose the condition of the observed relic density, 
$\mu_{eff}$ is almost determined by $m_{\chi}$. Therefore the mass 
spectrum of the Higgs does not change significantly. 
Two samples in Table \ref{tab:masses} satisfy the current experimental limits from LEPII and XENON10 and also explain the observed
relic abundance of dark matter. The mass of the gluino is $m_{\tilde{g}} \sim 200\  {\rm GeV}$. 
The lightness of the gluino is the characteristic feature of this scenario.

\TABLE[htbp]{
\caption{Mass spectra}
\begin{tabular}{|c|c c c c c c|}
\hline
\multicolumn{1}{|l|}{} & $m_0$ & $N_f$ & $d$ & $D_Y$ & $\lambda$ & $\tan\beta$ \\ \hline
input p1 & 200 & 1 & 2.2 & 4.84 & 0.69 & 2 \\ \hline
p2 & 200 & 1 & 3.5 & 8.74 & 0.69 & 2 \\ \hline
\end{tabular}
\begin{tabular}{|c|cccccccc|}
\hline
 & $|\mu_{eff}|$ & $m_{H_1^0}$ & $m_{H_2^0}$ & $m_{H_3^0}$ & $m_{A_1^0}$ & $m_{A_2^0}$ & $m_{H^{\pm}}$ & $m_{\chi_1^0}$ \\ \hline
output p1 & 264.2 & 127.2 & 328.8 & 460.1 & 285.5 & 487.8 & 433.8 & 34.2 \\ \hline
p2 & 270.6 & 127.5 & 313.5 & 590.0 & 286.6 & 601.6 & 582.7 & 34.3 \\ \hline \hline
 & $m_{\chi_2^0}$ & $m_{\chi_3^0}$ & $m_{\chi_4^0}$ & $m_{\chi_5^0}$ & $m_{\ti{g}}$ & $m_{\chi_1^{\pm}}$ & $m_{\chi_2^{\pm}}$ & $m_{\ti{\nu}_L}$ \\ \hline
p1 & 198.3 & 310.5 & 336.4 & 403.8 & 262.1 & 197.2 & 350.2 & 174.5 \\ \hline
p2 & 237.1 & 317.0 & 417.9 & 459.2 & 185.1 & 237.7 & 428.9 & 408.9 \\ \hline \hline
 & $m_{\ti{\nu}_\tau}$ & $m_{\ti{e}_L}$ & $m_{\ti{e}_R}$ & $m_{\ti{\tau}_1}$ & $m_{\ti{\tau}_2}$ & $m_{\ti{u}_L}$ & $m_{\ti{u}_R}$ & $m_{\ti{t}_1}$ \\ \hline
p1 & 174.4 & 184.5 & 651.7 & 184.3 & 651.6 & 1231.7 & 991.9 & 829.5 \\ \hline
p2 & 408.8 & 413.1 & 925.1 & 413.1 & 925.1 & 1682.2 & 1347.1 & 1126.7 \\ \hline \hline
 & $m_{\ti{t}_2}$ & $m_{\ti{d}_L}$ & $m_{\ti{d}_R}$ & $m_{\ti{b}_1}$ & $m_{\ti{b}_2}$ & $\Omega h^2$ & $\sigma_p^{SI} ({\rm cm}^2)$ &  \\ \hline
p1 & 1177.4 & 1233.2 & 1169.1 & 1166.9 & 1170.1 & 0.111 & $3.3\times 10^{-44}$ & \multicolumn{1}{l|}{} \\ \hline
p2 & 1605.4 & 1683.3 & 1573.3 & 1573.1 & 1598.9 & 0.131 & $3.1\times 10^{-44}$ & \multicolumn{1}{l|}{} \\ \hline
\end{tabular}
\label{tab:masses}
}

\section{Conclusions}
We investigated the phenomenology of the nMSSM with a Fayet-Iliopoulos D-term in the positively deflected anomaly mediation scenario.

In the deflected anomaly mediation scenario, the messenger sector is introduced.
We showed that the couplings between the nMSSM fields and the messenger sector fields are forbidden by the discrete symmetry,
and therefore the phenomenology at the weak scale is not affected by the detail of the
messenger sector.
We evaluated the soft breaking terms at the messenger scale without assuming small Yukawa couplings,
and showed that the contributions from Yukawa couplings are the same as those of anomaly mediation. 
The soft breaking parameters are determined by the deflection parameter $d$, 
the messenger scale and contributions from the Fayet-Iliopoulos D-term.

We also discussed the phenomenology of the nMSSM at the weak scale.
We found that electroweak symmetry breaking is successful, and moderate values of $\mu_{eff}$ are obtained.
The mass of the lightest Higgs is heavier than the LEP bound.
We also obtained sparticle mass spectra, and interestingly, the gluino is light.

We showed that the lightest neutralino is a good candidate for dark matter. 
The relic density explains the observed abundance of dark matter.
The spin-independent dark matter-proton scattering cross section 
satisfies the upper limit from XENON10 when we consider a small value of the strange quark content of the nucleon as 
indicated by a recent lattice calculation.
The cross section is large enough to be detected or excluded by next-generation experiments of direct detection. 

We consider this scenario phenomenologically viable. If the light gluino is discovered, it may imply
that SUSY breaking is mediated by supergravity and messengers,
and these two effects are comparable.

\section*{Acknowledgements}
We thank H. Ohki for useful discussion on the direct detection of dark matter and informing us of the nucleon sigma term.
We also thank M. Ibe for discussion on SUSY breaking effects of an intermediate threshold and DM-nucleon scattering cross section. 
We would like to thank K. Akina and T. Morozumi for careful reading of the manuscript.
We thank A. Masiero for discussion on symmetry breaking terms. 
We acknowledge M. Okawa, K. Ishikawa and T. Inagaki for support and encouragement.

\appendix

\section*{Appendix A: Explicit example of the positively-deflected 
 anomaly mediation scenario in the nMSSM}

In this appendix, we show that the positively deflected anomaly mediation 
scenario is achieved in the nMSSM.

Let us discuss the scalar potential of $X$. After rescaling the fields as $\hat{X}\phi \rightarrow \hat{X}$, the superpotential of $\hat{X}$ is
\begin{eqnarray}
W(\hat{X}) = \frac{m_X}{2} \hat{\phi} \hat{X}^2 . \label{eq:wx1}
\end{eqnarray}
We also consider tadpoles generated by the supergravity interaction.
\begin{eqnarray}
\Delta W(\hat{X}) = \mu_X^2 \hat{X} , \label{eq:wx2}
\end{eqnarray}
and
\begin{eqnarray}
\Delta \mathcal{L} = c_X \mu_X^2 X + h.c. \label{eq:wx3}
\end{eqnarray}
The tadpoles are expected to be much smaller than the plank scale owing to 
$Z_{7R'}$. 
 We assume that all couplings of positive mass dimension in the messenger sector are on the order of $F_\phi$
: $m_X \sim \mu_X \sim c_X \sim F_\phi$.

From eq. (\ref{eq:wx1}-\ref{eq:wx3}), we obtain the scalar potential for $X$,
\begin{eqnarray}
V(X) = \left|\mu_X^2 + m_X X \right|^2 - \left(c_X \mu_X^2 X - \frac{m_X}{2} F_\phi X^2 + h.c.\right) \ . 
\end{eqnarray}
We assume that CP is conserved for simplicity. The messenger scale, $X$, is determined by the stationary condition and 
\begin{eqnarray}
X = \frac{\mu_X^2 \left(c_X - m_X\right)}{m_X \left(m_X - F_\phi \right)} .
\end{eqnarray}
The minimum condition, $\frac{\partial^2 V}{\partial X^2} > 0$ leads to
$m_X > F_\phi$; therefore, the fermionic partner of $X$ is not the LSP. 
We now evaluate the deflection parameter:
\begin{eqnarray}
d F_\phi \equiv \frac{F_X}{X} - F_\phi = \frac{m_X(m_X-F_\phi)-(m_X+F_\phi)(m_X-c_X)}{m_X-c_X}.
\end{eqnarray}
When $c_X$ is close to $m_X$, $d$ is positive and large. 

\section*{Appendix B: Soft Breaking Masses and Trilinear Couplings}
In this appendix, we present derivations for the soft breaking mass terms and the scalar trilinear couplings.
We show that at the messenger scale, Yukawa contributions to the soft breaking terms are the same as those for anomaly mediation.

We write the anomalous dimension and beta-functions as
\begin{equation}
\gamma_i \equiv \frac{d \ln Z_i}{d \ln \mu} = \frac{1}{16\pi^2}\sum_{A} c_A^i g_A^2,  \label{eq:amdim}
\end{equation}
\begin{equation}
\frac{d g_a}{d \ln \mu} = -\frac{b_a}{16\pi^2} g_a^3 ,
\end{equation}
\begin{eqnarray}
\frac{d y}{d \ln \mu} = \frac{1}{16\pi^2}\left(d_a g_a^2 + d_y y^2 \right) y, 
\end{eqnarray}
where $g_A = \left\{g_1, g_2, g_3, y \right\}$. The coefficients of the gauge-coupling beta-function are $b_a-N_f$ above the messenger scale.

We obtain the wavefunction renormalization constant $Z_i$ by integrating out the anomalous dimension $\gamma_i$
in eq. (\ref{eq:amdim}) from $\mu$ to $\Lambda$.
\begin{eqnarray}
\ln Z_i(s,t) &=& \ln Z_i(\Lambda) - \frac{1}{16\pi^2}\sum_A c_A^i \left(\int_s^t g_A^2 (s',t) ds' +\int_t^{\ln \Lambda} g_A^2 (t') dt'\right)\ ,
\end{eqnarray}
where $s = \ln\mu$ and $t = \ln |X|$. To obtain soft breaking terms, we need to differentiate $\ln Z_i(s,t)$ with respect to $s$ and $t$.
For this purpose, we define
\begin{eqnarray}
I \equiv \int_s^t f(s',t) ds'+ \int_t^{\ln\Lambda} f(t') dt',
\end{eqnarray}
and differentiate this expression with respect to $s$ and $t$:
\begin{eqnarray}
\frac{\p I}{\p s} &=&  -f(s,t) \ , \nnn
\frac{\p I}{\p t} &=&  \int_s^t \frac{\p f(s',t)}{\p t} ds' + \left.f(s,t)\right|_{s=t} -f(t) \ ,\nnn
\frac{\p^2 I}{\p s^2} &=&  -\frac{\p f(s,t)}{\p s} \ ,\nnn
\frac{\p I}{\p s \p t} &=& -\frac{\p f(s,t)}{\p t} \ ,\nnn
\frac{\p I}{\p t^2} &=&  \int_s^t \frac{\p^2 f(s',t)}{\p t^2} ds' + \left.\frac{\p f(s,t)}{\p s}\right|_{s=t} + 
\left. 2\frac{\p f(s,t)}{\p t}\right|_{s=t} - \frac{\p f(t)}{\p t} \ . \label{eq:soft_help}
\end{eqnarray}
Using these formula, we obtain
\begin{eqnarray}
\left. \frac{\p \ln Z_i(s,t)}{\p s} \right|_{s=t} &=& \sum_A \frac{c_A^i g_A^2}{16\pi^2} \ ,\nnn
\left. \frac{\p \ln Z_i(s,t)}{\p t} \right|_{s=t} &=& 0 \ ,\nnn
\left. \frac{\p^2 \ln Z_i(s,t)}{\p s^2} \right|_{s=t} &=& \sum_A \frac{c_A^i g_A(t)}{8\pi^2} \left. \frac{\p g_A(s,t)}{\p s} \right|_{s=t} \ ,\nnn
\left. \frac{\p^2 \ln Z_i(s,t)}{\p s \p t} \right|_{s=t} &=& \sum_A \frac{c_A^i g_A(t)}{8\pi^2} \left. \frac{\p g_A(s,t)}{\p t} \right|_{s=t} \ ,\nnn
\left. \frac{\p^2 \ln Z_i(s,t)}{\p t^2} \right|_{s=t} &=& -\sum_A \frac{c_A^i g_A(t)}{8\pi^2} \left[\frac{\p g_A(s,t)}{\p s} + 2 \frac{\p g_A(s,t)}{\p t} - \frac{\p g_A(t)}{\p t} \right]_{s=t} . \label{eq:defl_zzz}
\end{eqnarray}
From eqs. (\ref{eq:formula}) and (\ref{eq:defl_zzz}), the scalar trilinear coupling is
\begin{eqnarray}
a_{ijk}(t) = -\frac{F_\phi}{2}\left[\gamma_i(t) + \gamma_j(t) + \gamma_k(t) \right] y_{ijk}.
\end{eqnarray}
Next we derive soft breaking masses. For the gauge coupling
\begin{eqnarray}
\frac{\p g_a(s,t)}{\p s} &=& -\frac{b_a}{16\pi^2} g_a^3 \ ,\nnn
\frac{\p g_a(s,t)}{\p t} &=& \frac{N_f}{16\pi^2} g_a^3 \ ,\nnn
\frac{\p g_a(t)}{\p t} &=& -\frac{b_a-N_f}{16\pi^2} g_a^3 . \label{eq:defl_gauge}
\end{eqnarray}
Therefore, from eqs. (\ref{eq:formula}), (\ref{eq:defl_zzz}) and (\ref{eq:defl_gauge}), the soft breaking mass is
\begin{eqnarray}
 \ti{m}_i^2(t) &=& \frac{|F_\phi|^2}{2(4\pi)^4} \sum_a c_a^i g_a^4(m)\left[b_a + N_f d(d+2) \right] ,
\end{eqnarray}
which agrees with \cite{dam-org}\cite{dam-pos}.
For Yukawa couplings,
\begin{eqnarray}
 y(s,t) &=& y(\Lambda) - \frac{1}{16\pi^2}\int_{s}^{t} ds' \left[d_a g_a^2(s',t) + d_y 
 y^2(s',t) \right] y(s',t) \nn \\
&& -  \frac{1}{16\pi^2}\int_{t}^{\ln \Lambda} dt' \left[d_a g_a^2(t') + d_y 
 y^2(t') \right] y(t') .
\end{eqnarray}
Using eq. (\ref{eq:soft_help}), we obtain
\begin{eqnarray}
\left. \frac{\p y(s,t)}{\p s} \right|_{s=t} &=& \frac{y(t)}{16\pi^2}\left[d_a g_a^2(s,t) + d_y 
 y^2(s,t) \right]_{s=t} = \beta_y(t) \ ,\nnn
\left. \frac{\p y(s,t)}{\p t} \right|_{s=t} &=& 0 \ ,\nnn
\frac{\p y(t)}{\p t} &=& \beta_y(t) .
\end{eqnarray}
Therefore Yukawa contributions to the soft breaking mass at the messenger scale are
\begin{eqnarray}
\delta \ti{m}_i^2 = -\frac{|F_\phi|^2}{4} \frac{c_y^i}{8\pi^2} y(t) \beta_y(t) ,
\end{eqnarray}
which are the same as those in anomaly mediation.

\section*{Appendix C: Anomalous Dimensions}
One-loop anomalous dimensions of $S$, $H_u$ and $H_d$ are
\begin{eqnarray}
\gamma_s &=& \frac{1}{16\pi^2} (-4\lambda^2) \ ,\nonumber \\
\gamma_{H_u} &=& \frac{1}{16\pi^2}(\frac{3}{5}g_1^2 + 3g_2^2 -6y_t^2 -2\lambda^2) \ ,\nonumber \\
\gamma_{H_d} &=& \frac{1}{16\pi^2}(\frac{3}{5}g_1^2 + 3g_2^2 -6y_b^2-2y_{\tau}^2-2\lambda^2) .
\end{eqnarray}
At the one-loop level, the anomalous dimensions of other fields are the same as those in the MSSM. $\beta_\lambda$ is
\begin{eqnarray}
\beta_\lambda = -\frac{\lambda}{2}\left(\gamma_S + \gamma_{H_u} + \gamma_{H_d} \right) .
\end{eqnarray}

\providecommand{\href}[2]{#2}\begingroup\raggedright\endgroup


\end{document}